\documentclass{emulateapj}
\slugcomment{{\sc Accepted to ApJS:} February 12, 2009} 

\shorttitle{Keck/HIRES spectrum of V838 Mon}
\shortauthors{Kami\'{n}ski et al.}

\begin{document}

\title{Keck/HIRES spectroscopy of V838 Monocerotis in October
  2005\footnotemark[*]}
\footnotetext[*]{
       The data presented herein were obtained at the W.M. Keck
  Observatory, which is operated as a scientific partnership among the
  California Institute of Technology, the University of California and
  the National Aeronautics and Space Administration. The Observatory
  was made possible by the generous financial support of the W.M. Keck
  Foundation.}


\author{T. Kami\'{n}ski\altaffilmark{1}, M. Schmidt\altaffilmark{1}, R. Tylenda\altaffilmark{1}, M. Konacki\altaffilmark{1,2}, and
  M. Gromadzki\altaffilmark{1}} 

\altaffiltext{1}{Department of Astrophysics, 
       Nicolaus Copernicus Astronomical Center, 		  
       Rabia\'{n}ska 8,
       87-100 Toru\'{n}, Poland}
       
\altaffiltext{2}{Astronomical Observatory, 
       A. Mickiewicz University, 
       S{\l}oneczna 36, 
       60-286 Pozna\'{n}, Poland}
        
\email{tomkam@ncac.torun.pl}



\begin{abstract}
V838~Mon erupted at the beginning of 2002 becoming an extremely
 luminous star with $L \simeq 10^6$~L$_{\sun}$. Among various scenarios
 proposed to explain the nature of the outburst the most promising is
 a stellar merger event. In this paper we investigate the
 observational properties of the star and its surroundings in 
  the post outburst phase. 
We have obtained a high resolution optical spectrum of V838~Mon 
in October 2005 using the Keck~I telescope. 
We have identified numerous atomic features and
  molecular bands present in the spectrum and provided an atlas of
  those features. In order to improve the spectrum
  interpretation we have performed simple modeling of the molecular bands.
Our analysis indicates that the spectrum is dominated by molecular
absorption features arising in photospheric regions with temperatures of
  $\sim$2400~K and in colder outer layers, where the temperature decreases
to $\sim$500~K. A number of resonance lines of neutral alkali metals
  are observed to show P-Cyg profiles. Particularly interesting are 
numerous prominent emission
  lines of [Fe~{\small II}]. All of them show practically the same profile,
which can be well described by a Lorentzian profile. 
In the blue part of the spectrum photospheric signatures of the B-type
  companion are easily seen. We have fitted the observed spectrum with a
synthetic one and the obtained parameters are consistent with the B3V type.
We have also estimated radial and rotational velocities of
  the companion. 
\end{abstract}

       

\section{Introduction \label{intro}}

The eruption of V838 Mon was discovered in the beginning of January~2002.
Initially thought to be a nova, the object appeared unusual and enigmatic in
its nature. The eruption, as observed in the optical, lasted about three
months \citep{muna02,kimes02,crause03}. After developing an A-F supergiant spectrum 
at the maximum at the beginning of February~2002, the object evolved to
lower effective temperatures and in April~2002 it practically disappeared
from the optical, remaining very bright in the infrared. At the same time a
B3V companion to the erupted object was discovered in the optical \citep{mundes}. 
A detailed analysis
of the evolution of the object in the outburst and decline can be found,
e.g. in \cite{tyl05}.

Several mechanisms have
been proposed to explain the eruption of V838~Mon, 
including an unusual nova 
\citep{it92}, a late He-shell flash \citep{law05}, and a stellar
merger \citep{soktyl03}.
They have critically been discussed in \cite{tylsok06}.
These authors conclude that the only mechanism that
can satisfactorily account for the observational data is a collision and
merger of a low-mass pre-main-sequence star with a $\sim 8\,M_\odot$
main-sequence star.

In October/November 2006 the B3V companion
significantly faded and a strong H$\alpha$ emission appeared in the spectrum
of V838~Mon \citep{gor06,bond06}. Late in 2004 an emission-line spectrum,
composed mainly of [\ion{Fe}{2}] lines, started to develop \citep{barsuk06}, 
reaching its maximum around the 2006 eclipse-like event \citep{mun07}.

In the present paper we present and discuss a high resolution spectrum 
of V838~Mon acquired with the Keck~I telescope in October 2005. 
At that time the [\ion{Fe}{2}] emission-line spectrum was already well developed
and the object was a year before the eclipse of the B3V companion.

\section{Observations and data reduction \label{obsred}}

A high resolution spectrum of V838~Mon was obtained on 2005 October 13 UT with 
the High Resolution Echelle Spectrometer (HIRES, \citealp{vogt}) attached to 
the Keck~I telescope. We used the C4 decker, which provides a slit with dimensions 
1$\farcs$148 $\times$ 3$\farcs$5. The resultant resolving power was of 
$R\simeq 34\,000$. The HIRES instrument makes use of a 3-chip mosaic as a detector. 
We obtained two exposures of the object with a total time of 960$\,$s.  

Data reduction was performed with the IRAF\footnote{IRAF is distributed by 
the National Optical Astronomy Observatories, which are operated by 
the Association of Universities for Research in Astronomy, Inc., under 
cooperative agreement with the National Science Foundation.} standard
procedures \citep{iraf}. 
Data were debiased and flat-fielded using the {\it ccdproc} task. Cosmic ray events 
were removed from the CCD frames with a method described in \cite{dokkum}. 
After a careful check of the removal we are confident that the procedure did not 
introduce any spurious effects to the spectra. Order tracing and extraction were 
done using the tasks in the {\it echelle} package. The orders corresponding to the 
bluest part of the spectra were underexposed and we were not able to extract them, 
even with a pinhole trace as a reference. The wavelength calibration was based 
on ThAr lamp observations and it is accurate to within 0.003$\,$\AA. The two extracted 
echelle spectra were averaged giving a spectrum with a total wavelength coverage 
$3720\,-\,7962\,$\AA. In the range there are, however, numerous gaps caused by 
the inter-chip spaces on the mosaic and the orders that do not overlap.

Using an extracted pinhole frame we corrected all the orders of the echelle
spectrum for a blaze effect. Inspection of the overlapping and blaze-corrected orders 
revealed that, although correctly flattened, they were systematically tilted in 
such a way that the long-wavelength end of an order was always higher on a relative 
intensity scale than the short-wavelength end of the neighboring order. 
The discrepancy was of 15\% of the average intensity of the order. We reduced 
the tilt by multiplying each order by a linear function 
(i.e. with $1$ in the middle of an order range, and $1\pm 0.075$ at the edges). 
All the overlapping orders were then safely merged and, subsequently, we got 
a spectrum composed of three parts: {\it (i)}  completely covered wavelength range 
over $3720\,-\,4883\,$\AA, {\it (ii)} wide range over $4951\,-\,6469\,$\AA\ 
with two narrow gaps at $6308.5\,-\,6310\,$\AA\ and $6401\,-\,6424.5\,$\AA, 
and {\it (iii)} a part of the spectrum over the range $6544\,-\,7962\,$\AA\ 
with 10 non-overlapping orders. The three pieces we call hereafter parts blue, green and
red, respectively. 

Since no spectrophotometric standard stars were observed during our observing run
at Keck~I, 
we decided to flux calibrate the HIRES spectrum on a base of the method described 
in \cite{suzuki}, namely by using a flux calibrated spectrum of V838~Mon obtained 
elsewhere with a lower spectral resolution. Such a low resolution spectrum was 
acquired on 2005~November~22 with the Grating Spectrograph with SITe mounted at 
the 1.9~m Radcliffe telescope at the South African Astronomical Observatory (SAAO). 
We made use of a grating \#7 with 300 lines mm$^{-1}$, and a slit with a projected 
width  1$\farcs$5. The last figure was comparable to seeing conditions during 
the observation ($\sim$1$\farcs$4). The total duration of the exposure was 1000$\,$s. 
Three spectrophotometric standard stars, i.e. 
LTT9239, EG21, and LTT3218, 
were observed during the night. This allowed us to calibrate 
the low resolution spectrum in absolute flux units, while the wavelength calibration 
was performed with ThAr reference lamp spectra. All the data reduction and 
calibrations were carried out with standard IRAF procedures. The extracted and 
flux calibrated spectrum covers the range $3817\,-\,7235\,$\AA\ with 
a resolving power of $R\simeq1\,000$.

The flux calibration of the HIRES spectrum was performed as follows. 
The high and low resolution spectra were corrected to the heliocentric rest frame 
and they were cross-correlated in order to find a relative shift between them 
in the wavelength domain. We found a shift of 5.5$\,$\AA\ (note that the value is 
close to the spectral resolution in the SAAO observations), which was eliminated 
by displacing the low resolution spectrum. Next, the HIRES spectrum was smoothed 
to the resolution of the observations at SAAO and the spectra were divided one by the other. 
Rather than to calibrate each single order of the echelle spectrum, as it was done 
in \cite{suzuki}, we decided to calibrate separately only 
the three mentioned above parts of the merged spectrum. 
To each of the pieces a low order polynomial was 
fitted giving a conversion ratio \citep[CR, cf.][]{suzuki}. 
Since the SAAO spectrum 
has a narrower wavelength range than the high resolution one, we extrapolated 
the CRs to the full ranges of the blue and red parts. Finally, the HIRES spectrum was 
put on the absolute flux scale by dividing its parts by the appropriate CRs. 
We note that the flux calibration of our final spectrum is uncertain,
particularly in the ranges not covered by 
the SAAO observations and in the regions where strong atmospheric molecular bands 
are present (the object was observed at unequal air masses in the different 
facilities).      

Although the low resolution spectrum was obtained about six weeks after 
the HIRES observations, we are quite confident that the spectrum of V838~Mon 
did not change significantly during this period in the observed spectral range. 
A comparison of the spectra smoothed to the same resolution shows that there are 
no significant discrepancies which would affect the flux calibration. 
Moreover, from the photometric behavior of the object, illustrated e.g. on 
the V.~Goranskij's web page\footnote{\tt http://jet.sao.ru/$\sim$goray/v838mon.htm}, 
one can see that the assumption of a constant flux in B and V bands during 
the period is justified.

The flux calibrated spectrum was then corrected for the interstellar reddening 
using the IRAF's task {\it deredden} with $E_{B-V} = 0.90$
\citep{tyl05} 
and with a standard ratio of total to selective extinction
$R_V=3.1$. 

%
\section{The spectrum \label{results}}

\begin{figure*}
\centering
   \includegraphics[width=14cm]{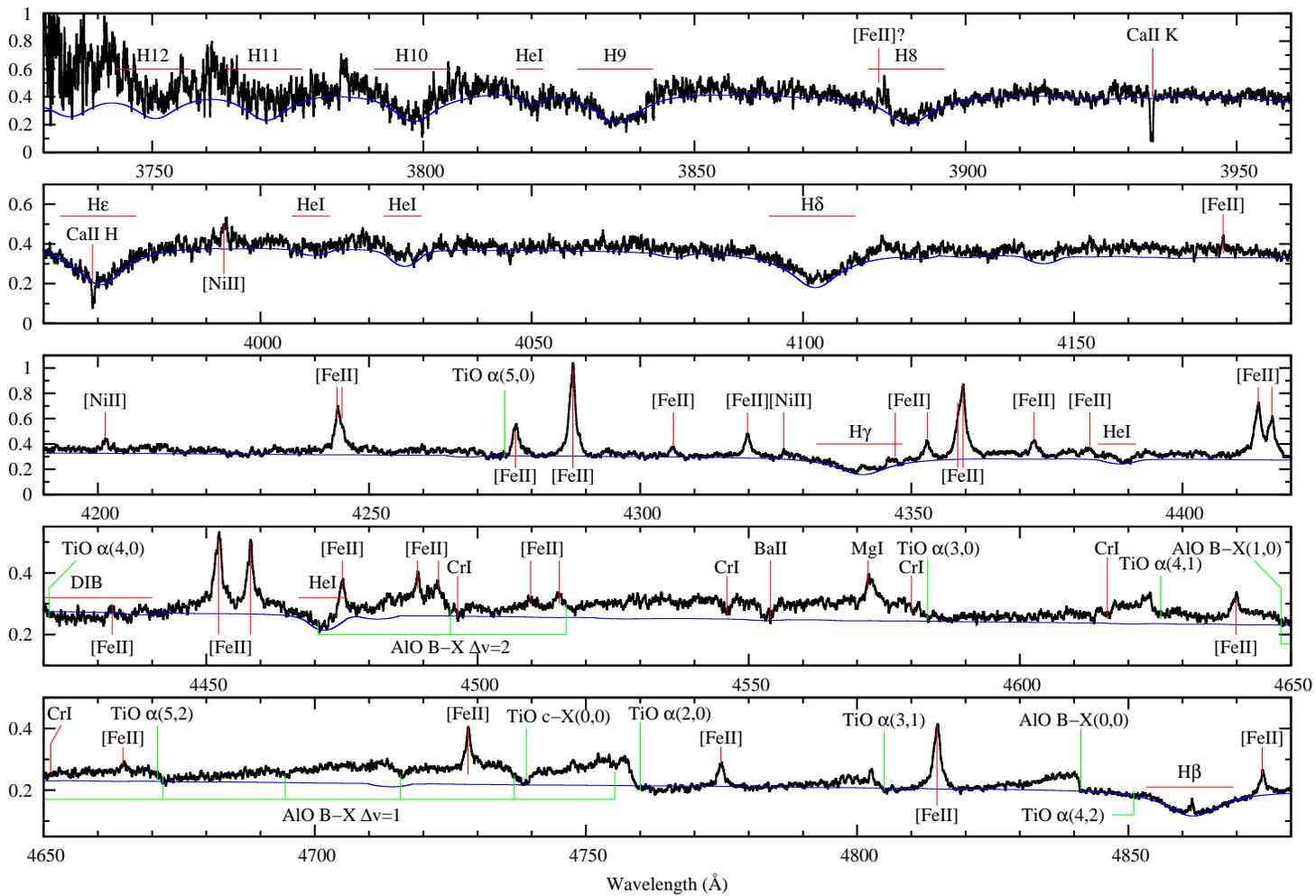}
   \caption{Blue part of the Keck spectrum of V838 Mon obtained in 
     October~2005. The spectrum was smoothed from the original resolution with
     boxcar~13. A synthetic spectrum of a
     B3V star is shown for comparison (blue line). Identified atomic spectral
     features are indicated by red markers, while molecular bandheads
     are assigned with green markers. The ordinate units are
     $10^{-13}$~erg~s$^{-1}$~cm$^{-2}$~\AA$^{-1}$.} 
     \label{specB}
\end{figure*}

\begin{figure*}
\centering
   \includegraphics[width=14cm]{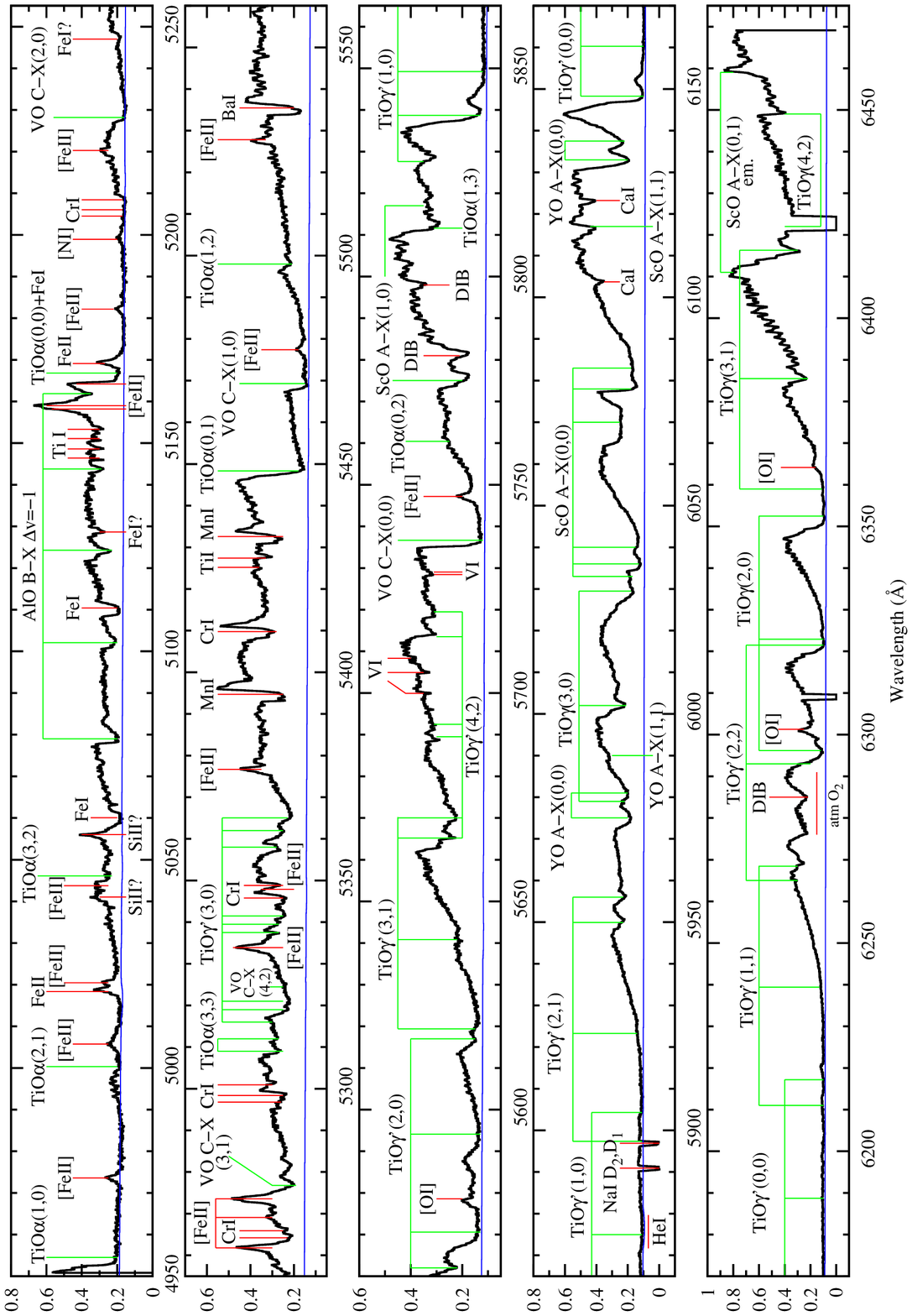}
   \caption{Same as in Fig.~\ref{specB} but for the green part of 
   the Keck spectrum of V838 Mon.}
   \label{specG}
\end{figure*}

\begin{figure*}
\centering
   \includegraphics[width=14cm]{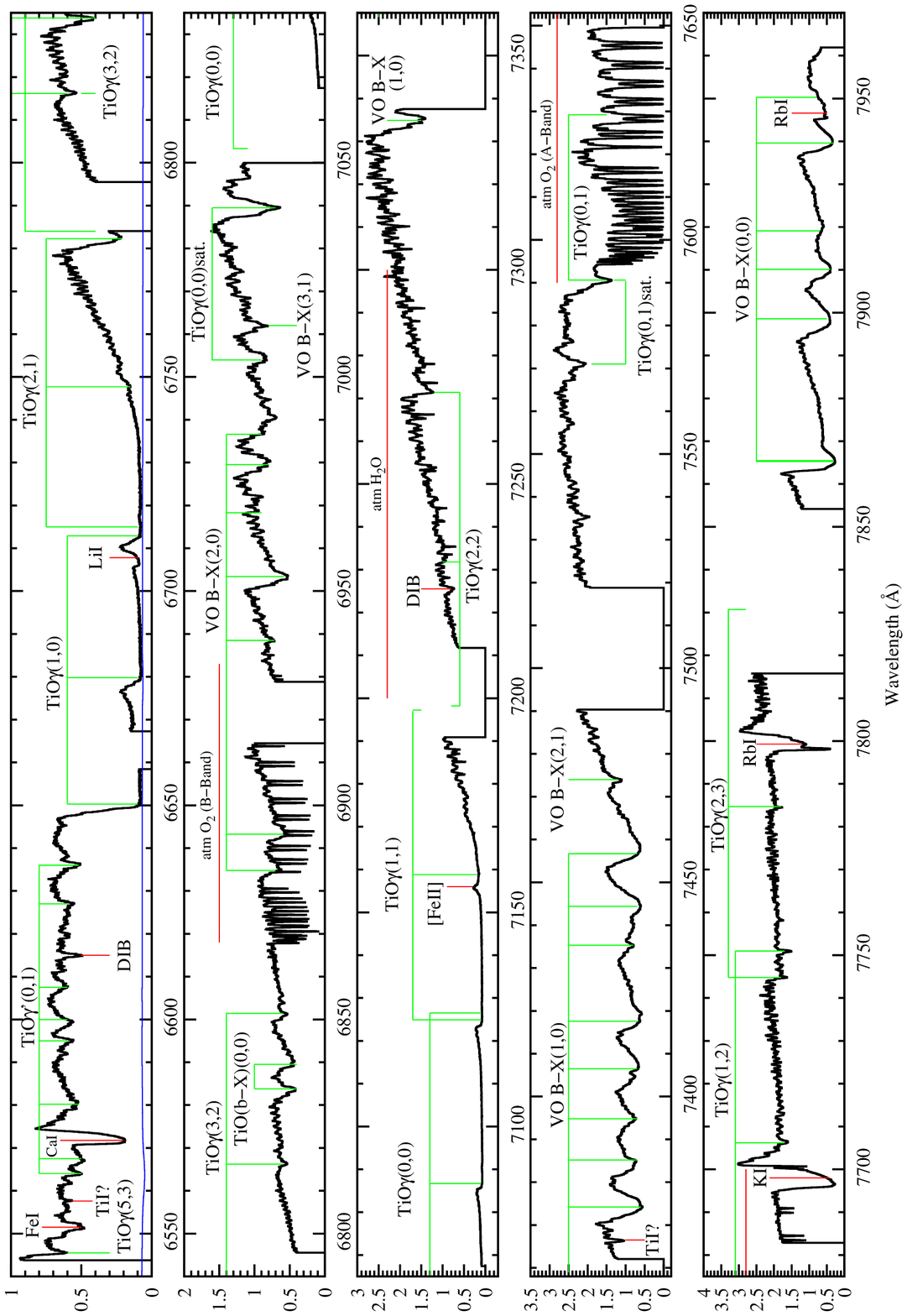}
     \caption{Same as in Fig~\ref{specB} but for the red part of 
the Keck spectrum of V838 Mon.}
     \label{specR}
\end{figure*}

The flux-calibrated and dereddened spectrum of V838~Mon is presented in
Figs. \ref{specB}--\ref{specR}. For a comparison a synthetic spectrum of a 
B-type star (see Sect. \ref{hotstar}) is overploted with the blue line.
Most of identified atomic and molecular features are
indicated in the figures. Wavelengths are given  in the heliocentric
rest frame and the units of the flux are 
$10^{-13}$~erg~s$^{-1}$~cm$^{-2}$~\AA$^{-1}$.
 

Our observations reveal an unusual and very complex spectrum of
V838~Mon. It contains contributions from several physically different
environments. In its blue part a photospheric spectrum of the B3V
companion can be clearly seen, while the green and red parts are
dominated by molecular absorptions that severly affect the flux of a very cold
supergiant. 
A particularly striking feature of the presented spectrum is the presence of prominent
emissions of [\ion{Fe}{2}]. Finally numerous, mostly resonance, lines showing
P-Cyg profiles as well as a few pure absorptions of neutral atoms 
can be easily found.

In the case of such a complex spectrum, detailed identification
 of all the features is a difficult task. The identification
 procedures and their results are described in the following
 sections. We also present results of basic measurements, e.g. 
radial velocities of
 different line systems and molecular bands.   

\section{Atomic lines}
The atomic line identification was based mainly on the Atomic Spectra
Database Lines Form\footnote{\tt 
http://physics.nist.gov/PhysRefData/ASD/lines\_form.html} (provided by
the National 
Institute of Standards and Technology, NIST), the Atomic Line List version 2.04 by
van Hoof\footnote{\tt http://www.pa.uky.edu/\-$\sim$peter/\-atomic/},
and on the multiplet tables by \cite{moore}. 

All the identified atomic emission and absorption features are listed in Tables
\ref{atom_em_tab} and \ref{atom_ab_tab} (respectively) in order of
their laboratory wavelengths. Columns (1)--(3) of the tables contain:
laboratory wavelength, ion name, and multiplet number from
\cite{moore}.  Column (4) in Table~\ref{atom_em_tab} gives
integrated (dereddened) fluxes of the emission lines (for which a reliable
measurement was possible). In the same column of
Table~\ref{atom_ab_tab} the profile type (pure absorption or P-Cyg) is
specified. Notes on the features are given in the last 
column of the tables.

\subsection{Emission lines  \label{em_lines}}

\subsubsection{Forbidden lines \label{forbid_lines}}
A great majority of the emission features seen in the spectrum are forbidden lines
of \ion{Fe}{2}. They are remarkably numerous. We have identified 49
lines from 11 different multiplets. All of them arise from levels with
energies (above the ground level) between 2.0 and 3.4~eV. Most of them are very strong and often
blended with other spectral features. In
addition to these forbidden transitions, we have also found two permitted lines of
\ion{Fe}{2} at 5018~\AA\ and 5169~\AA\ (a third line from the same multiplet falls in 
a gap between two chips). 

\begin{figure}
\centering
\includegraphics[scale=0.35,angle=270]{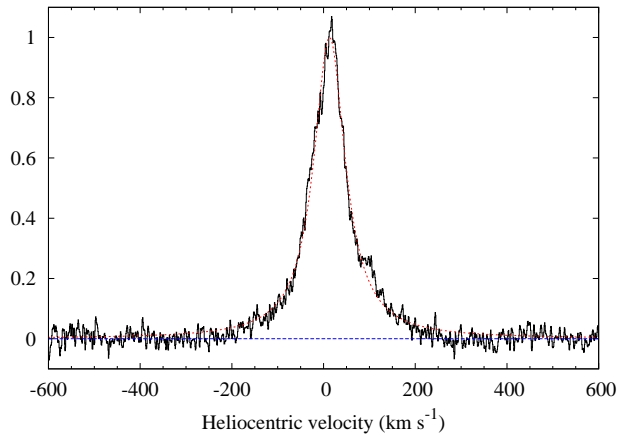}
  \caption{Profile of the [\ion{Fe}{2}] lines.
{\it Black solid line:} 
the profile averaged out from 7 lines (see the text for more
details).   
{\it Red dotted line:} 
the best least-squares fit of a Lorentzian profile to the average profile 
(FWHM of 83.6~km\,s$^{-1}$, center at
13.0~km\,s$^{-1}$).
{\it Blue dashed line:} local continuum.}
  \label{FeII_fig}
\end{figure}

All the unblended [\ion{Fe}{2}]
emissions show practically the same but rather unusual shape. It can be
well fitted with a
theoretical Lorentzian profile, as can be seen from Fig.~\ref{FeII_fig}.
The profile shown in this figure is a mean shape obtained from
seven strongest and ``clean" [\ion{Fe}{2}] lines. 
When averaging, the
normalized profiles of individual lines were weighted
according to the rms$^2$ values of the local noise in the spectrum.
From 30 [\ion{Fe}{2}] lines, for
which we successfully fitted a Lorentzian profile using IRAF's {\it splot},
we have obtained a typical full width at half
maximum\footnote{All line widths given in this paper are not corrected
                 for the instrumental profile, which has an intrinsic 
                 width of about 9~km~s$^{-1}$.} 
(FWHM) of 76.3$\pm$17.7~km~s$^{-1}$ (median and standard deviation). 
All the lines have practically the same radial velocity.
Measurements performed
with the IRAF's {\it rvidlines} task gave a heliocentric radial
velocity of $V_h$=$13.3\pm0.7$~km~s$^{-1}$ (mean and standard deviation). 

Other forbidden lines seen in our spectrum include those of 
[\ion{O}{1}] 5577~\AA, 6300~\AA,
6364~\AA\  and [\ion{N}{1}] 5198~\AA.
These are rather weak features, observed in spectral regions strongly
contaminated with molecular bands. Their measurements are therefore
uncertain. The measured FWHM and
radial velocity of the [\ion{O}{1}] lines are $80\pm30$~km~s$^{-1}$
and $40\pm10$~km~s$^{-1}$, respectively.
The [\ion{N}{1}] line has a radial velocity
of about 80~km~s$^{-1}$. 
 
It should be noted that some of the
emission lines listed in Table~\ref{atom_em_tab} have uncertain identification. 
Two strong features at about 5041~\AA\ and 5056~\AA\ were identified as
\ion{Si}{2}, but other identifications are possible. Three
lines recognized as [\ion{Ni}{2}] 3993~\AA,
4201~\AA, and 4326~\AA, are weak, which makes their identification
uncertain. However, the ionization potential of Si and Ni being similar to that 
of Fe and low excitation energies of the
proposed multiplets make the identification probable.

\subsubsection{Balmer lines  \label{balmer_lines}}

\begin{figure}
\centering
\includegraphics[scale=.4]{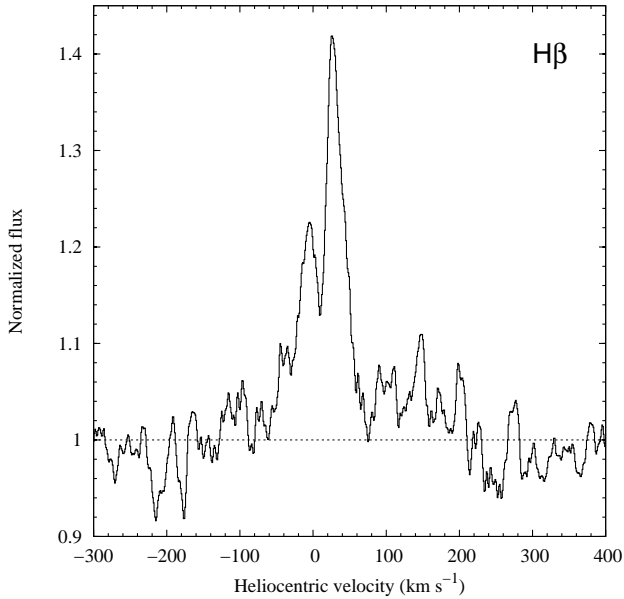}
  \caption{Emission feature extracted from the H$\beta$ photospheric
  absorption line of the B3V companion. The spectrum was
  significantly smoothed from the original resolution via boxcar with
  a box size of 15 pixel.}
  \label{Hemission}
\end{figure}

Inside the cores of the photospheric Balmer lines of the B-type companion
weak emission features can be found. Their
presence is evident when one compares the observed photospheric
profiles with a synthetic spectrum of an early B-type star (see
Fig. \ref{specB}). The following discussion is limited to H$\beta$ only, as 
the emission feature is most evident in this line.
In order to better characterize the emission we have
subtracted the underlying stellar absorption. The procedure was as
follows. To the wings of the absorption line we fitted a profile in
the form:   
\begin{equation}
\label{profil}
\phi(\lambda)=d\exp\{[-a(\lambda \pm \lambda_c)^b+c]^{-1}\}
\end{equation} 
with $a$, $b$, $c$ and $d$ as free parameters. The
central wavelength, $\lambda_c$, was taken as the laboratory
wavelength of the H$\beta$ line corrected for the radial velocity of
the B3V star (see Sect.~\ref{hotstar}). Eq.~(\ref{profil}) is based on an
empirical profile given in \cite{chauville}. 
However, for our purpose it was
generalized to treat also the local continuum level as a free
parameter. The fitting procedure used the least-squares method and it 
was applied to all the data points in the H$\beta$ wing [i.e., not only to three
points, as in \cite{chauville}]. The core region
dominated by emissions was, obviously, excluded from the fit. The
extracted emission feature is shown in Fig.~\ref{Hemission}. It can be
interpreted as a single emission with an absorption superimposed on it.
In this case, fitting Lorentzian profiles gives the emission component with a FWHM
of $\sim$42~km~s$^{-1}$ at a radial velocity of $\sim$14~km~s$^{-1}$, while
the absorption component appears at $\sim$10~km~s$^{-1}$ and has a FWHM of
$\sim$24~km~s$^{-1}$. Note that the parameters of the emission component in this
case are close to those of the [\ion{Fe}{2}] lines.
Alternatively, if the observed profile is fitted with two
emission components, they have radial velocities of $\sim -$10 and
$\sim$28.8~km~s$^{-1}$, and FWHMs of $\sim$30 and $\sim$24~km~s$^{-1}$,
respectively.

\subsection{P-Cyg features}\label{pcyg}
\begin{figure}
\centering
\includegraphics[scale=.97]{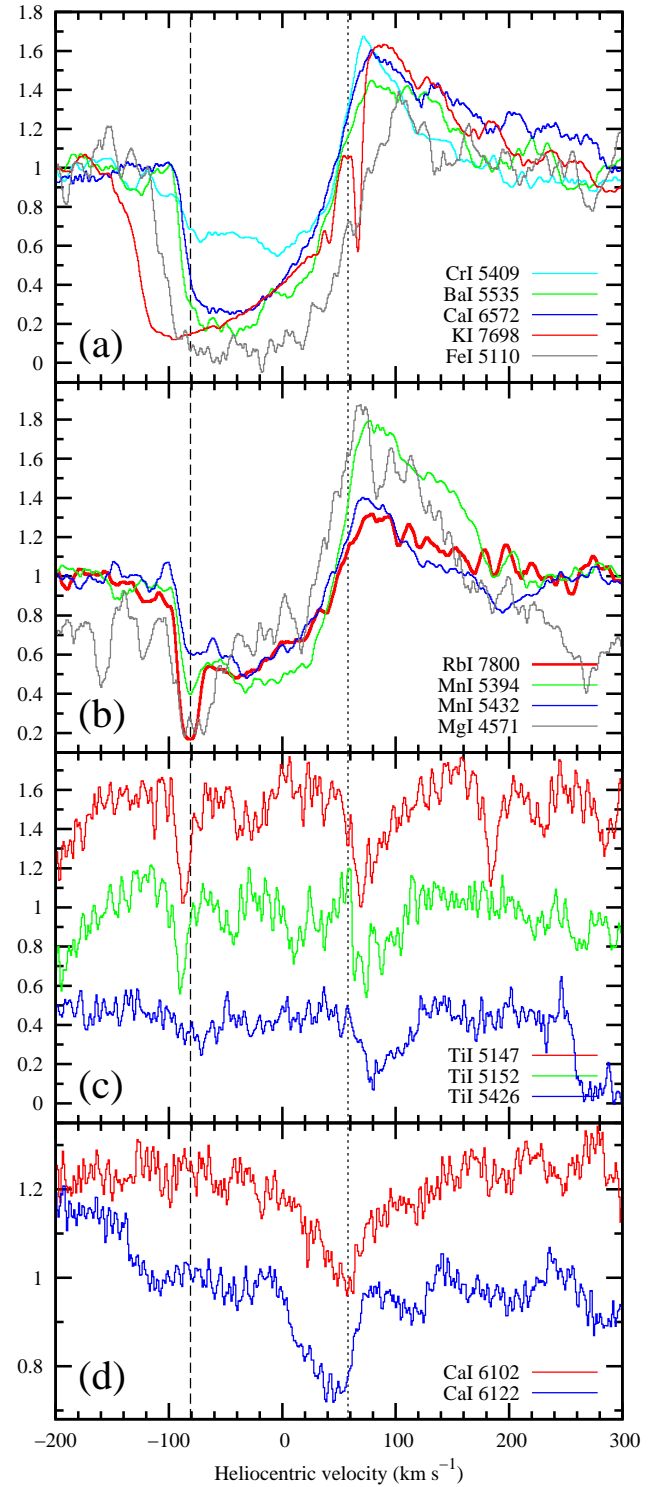}
\caption{Sample of atomic spectral features. Before the regions were
           individually normalized to the local continuum, the flux from the
           B3V star had been subtracted. 
           {\bf (a)} Examples of the most prominent
           P-Cyg lines. The spectra of \ion{Cr}{1}, \ion{Ba}{1} and
           \ion{Fe}{1}, were smoothed via boxcar 9. The \ion{K}{1}
           profile is contaminated by the ISM absorptions.
           {\bf (b)} P-Cyg profiles with a narrow absorption component
           (NAC). The normalized fluxes of
           the \ion{Mg}{1} were divided 
           by 1.7 and the spectrum was smoothed via boxcar 13. The
           \ion{Mn}{1} lines were smoothed with boxcar 9.  
           {\bf (c)} Double absorption components of
           three \ion{Ti}{1} transitions; the upper and lower plots
           were vertically shifted by $\pm$0.55. {\bf (d)} Two absorption
           lines of \ion{Ca}{1}. The vertical
           lines along the panels mark the velocity of the NAC
           (dashed) and of the source of the outflow (dotted) as estimated
           from molecular bands (see Sect.~\ref{tio}). 
           }
\label{pcyg_fig}
\end{figure}

The spectrum displays a number of strong resonance lines showing
P-Cyg profiles. The most prominent ones are those of \ion{Mn}{1}
5395~\AA\ and 5433~\AA, \ion{Cr}{1} 5410~\AA, \ion{Ba}{1} 5535~\AA,
\ion{Ca}{1} 6573~\AA, \ion{K}{1} 7699~\AA, \ion{Rb}{1} 7800~\AA\ (only
the \ion{Cr}{1} line is not a resonant one). They
are shown in Fig.~\ref{pcyg_fig}. Other lines showing P-Cyg signatures,
although not as notable as in those listed above, are \ion{Mg}{1}
4571~\AA, \ion{Fe}{1} 5110~\AA, and possibly also a few weak lines of
\ion{Cr}{1}.  

Most of the P-Cyg profiles have a deep absorption component extending
from $\sim$50~km~s$^{-1}$ (as a cross-over from emission to
absorption) to $\sim -$95~km~s$^{-1}$. The most extreme exception is
the \ion{K}{1} line,
which shows blueshifted absorption down to
$\sim -$145~km~s$^{-1}$. Emission components 
peak at $\sim$80~km~s$^{-1}$. 

The shapes of the absorption components indicate inhomogeneities in the
outflowing matter. Particularly in the profiles of \ion{Rb}{1} and
\ion{Mn}{1} one can clearly see an additional narrow absorption
component (NAC) centered at $-$82~km~s$^{-1}$ [see panel (b) in
Fig. \ref{pcyg_fig}]. The measured FWHM of the narrow feature inside the
\ion{Rb}{1} 7800~\AA\ line is 19.3~km~s$^{-1}$. 

As noted above, the \ion{Mg}{1} 4571~\AA\ line shows only weak
signatures of a P-Cyg profile. Indeed, it is dominated by a slightly 
asymmetric emission and exhibits only a very weak absorption
at approximately the same velocity as the NAC.

The profile of \ion{Fe}{1} 5110~\AA\ is also exceptional. Its
absorption component is so deep that it removes all the light from
the cold star and 
reaches the flux level of the B3V companion (see top panel in
Fig.~\ref{specG}). In addition, it is considerably broader than most
of the other P-Cyg absorptions and extends from 
$\sim$80~km~s$^{-1}$ to $\sim -$120~km~s$^{-1}$. 
Most probably the $\lambda$5110 absorption is saturated. 
Other \ion{Fe}{1} lines identified in the spectrum appear as absorptions
only, without any signs of emission components (see Sect.~\ref{abs}).           

As Table~\ref{atom_ab_tab} indicates, almost all the abundant alkali
metals show P-Cyg profiles. Thus, one would
expect that it should also be the case of the \ion{Li}{1} doublet at 6708~\AA.
However, in the raw spectrum no clear
emission component can be 
seen close to the clear absorption of \ion{Li}{1}. The doublet falls
in a region that is dominated by a strong absorption band of TiO and it is difficult to
define an underlying molecular baseline. Moreover, the doublet resides very close
to the $\gamma$ (1,0) bandhead of TiO and most probably a part of the
emission component of \ion{Li}{1} is absorbed by the saturated
molecular band. 
A similar situation also occurs for the \ion{Rb}{1} 7948~\AA\ line. Its
profile is strongly affected by the (B$-$X) (0,0) absorption band of
VO and the expected emission component cannot be seen on the raw
spectrogram.

\subsection{Absorption lines}\label{abs}
Another group of identified lines are pure absorptions. Beside the
interstellar features (see Sect.~\ref{ism}) and photospheric lines
of a B-type star (see Sect.~\ref{hotstar}), absorptions of
\ion{Fe}{1} 5060~\AA, 5128\AA, 5247~\AA, 6551~\AA, \ion{Ca}{1}
6102~\AA\ and 6122~\AA, and numerous
\ion{Cr}{1} lines were found. Some of these lines are in the same
multiplet with a line that exhibits a prominent P-Cyg profile. Thus,
it is quite possible that in a more sensitive observations the lines,
classified here as absorptions, would appear as P-Cyg profiles. 
Among the absorptions the strongest is the \ion{Fe}{1}
5060~\AA\ line, which, similarly to the P-Cyg absorption 
component of \ion{Fe}{1} at 5110~\AA\ (see Sect.~\ref{pcyg}), appears as
a very broad feature and, probably, is saturated. All the absorptions,
except the $\lambda$5060 line, are narrower than the absorption components
in the P-Cyg profiles
and always fall in the velocity range defined by
P-Cyg absorptions, indicating that they originate in the same parts
of the outflowing gas.      

The absorption lines of \ion{Ti}{1} found 
at 5147~\AA, 5152~\AA, and 5426~\AA\ are of special interest. They arise
from levels with energies of 2.3$-$2.4~eV. In our spectrum the lines
consist of (at least) two well separated components, what is clearly 
seen in panel (c) of Fig.~\ref{pcyg_fig}. The redshifted
absorption component appears at $\sim$75~km~s$^{-1}$, close to
the peaks in the P-Cyg emissions. The blueshifted sharp
feature, seen in the 5147~\AA\ and 5152~\AA\ lines, appears at a radial 
velocity of $\sim -$90~km~s$^{-1}$,
which suggests that it can be associated with portions of
the fast wind seen in the most blueshifted P-Cyg absorptions
(excluding \ion{K}{1}). Note also that the velocity of
this short-wavelength component is also very close to
the velocity of the NAC seen in some P-Cyg profiles 
(see Fig.~\ref{pcyg_fig}).
As can be seen from panel (c) in Fig.~\ref{pcyg_fig},
the two absorption components of the \ion{Ti}{1} lines have different shapes. 
While the blushifted component has a very narrow Gaussian profile with a 
FWHM of $\sim$13~km~s$^{-1}$, the redshifted feature is
broad, with a FWHM of $\sim$40~km~s$^{-1}$, and its profile is
asymmetric with a steep blue edge and an extended red wing. 

A prominent absorption feature is present at $\sim$7366~\AA,
which we have identified as another \ion{Ti}{1} line at 7364~\AA. 
The feature would correspond to the blueshifted
component in the discussed above \ion{Ti}{1} lines. 
Note that the feature has also a profile similar 
to the corresponding absorptions in the other \ion{Ti}{1} lines. 
A possible redshifted component cannot be seen,
because it falls in a region between the echelle orders. 

We have also identified two weak absorption lines of \ion{Ca}{1} at
6103~\AA\, and 6122~\AA. They are shown in panel (d) of
Fig.~\ref{pcyg_fig}. These lines are broad (FWHM$\simeq$60\,km\,s$^{-1}$) and
asymmetric. Contrary to the \ion{Ti}{1} lines, they have an extended
blue wing and a sharp red edge. Since these \ion{Ca}{1} lines
arise from a rather high energy level (3.91\,eV above the ground level), they are
expected to be formed close to the photosphere of the red supergiant.    
 
\section{Molecular bands}\label{molecules}

The complexity of the spectrum makes a direct identification of molecular
bands very difficult. For identification purposes, we have
performed a series of simulations of molecular spectra. First attempts
showed that a model capable of reasonably reproducing the observed
molecular features should include, at least, two components,
i.e., a stellar photosphere of a cool supergiant and an absorbing
outflowing material with an excitation temperature significantly lower
than the stellar temperature. 

As a model of the stellar cold photosphere, we have used a synthetic
spectrum generated from a model atmosphere from the NextGen grid\footnote{
\tt  ftp://ftp.hs.uni-hamburg.de/pub/outgoing/phoenix/NG-giant} 
\citep{haus99}. These models include TiO opacities, what is crucial in the case
of the analyzed spectrum. We chose a model atmosphere with T$_{\rm
  eff}$\,=\,2400\,K, and with the lowest gravity available within the
grid, i.e. $\log{g} = 0.0$. For these parameters there is only a NextGen 
model with solar metallicity, what is probably close to the true
metallicity of V838~Mon \citep{kipperskoda}.  

Our choice on $T_{\rm eff}$ is a result of initial modeling of the
observed spectrum, where we used atmosphere models for $T_{\rm
  eff}\,=\,$2000, 2200, and 2400\,K. With the value of 2400\,K we have
obtained most satisfying fits to the observed band profiles from
higher rotational levels. 

\begin{figure}
 \centering
 \includegraphics[scale=0.4]{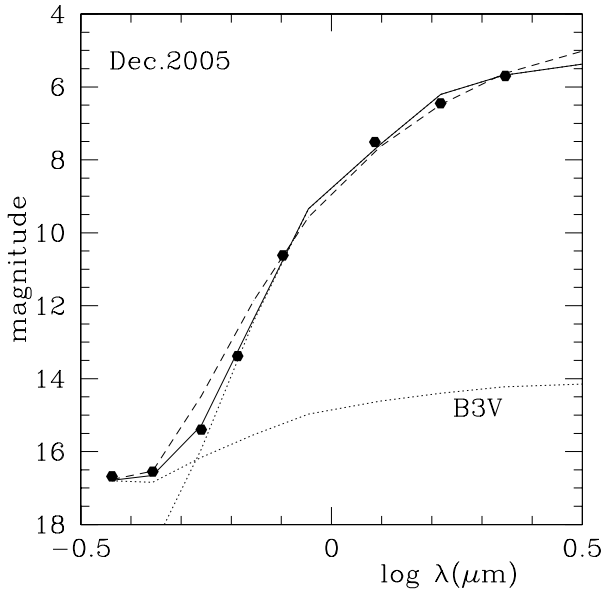}
 \caption{Model spectra fitted to the photometric ($UBVR_cI_cJHK'$ --
   full points) measurements of V838~Mon in December~2005. Dotted
   curves -- contributions from the supergiant and the B3V star. Full 
curve -- final fit (sum of the dotted ones). Dashed curve -- fit of a
blackbody to the $I_cJHK'$ measurements. See the text for more details.
} 
 \label{phot_1450}
\end{figure}

The effective temperature of the cold star can also be determined from 
photometric measurements. \cite{mun07}
measured $BVR_cI _c$ magnitudes of V838~Mon on 2005~Dec.~1, while
\cite{hen06} measured $UBVR_cI_c$ on 2005~Dec.~25 and $JHK'$ magnitudes
on Dec.~18. Thus the photometry was done $\sim$2~months after our
spectroscopy. Given very slow evolution of the object at that epoch we can
safely assume that the magnitudes did not change significantly between
October and December 2005. Applying the same procedure to the above
photometric results as that used in \cite{tyl05}, i.e. fitting a standard   
supergiant photometric spectrum with $E_{B-V}\,=\,0.9$ and taking into account
a contribution from a standard B3V star, we have obtained a spectral type of
M6--7 and $T_{\rm eff} \simeq 2480$~K. Adopting $R\,=\,3.1$ and a
distance of 8~kpc \citep[as in][]{tyl05} the effective
radius and luminosity of V838~Mon are $\sim 840\,R_\odot$ and $\sim 2.4
\times 10^4\,L_\odot$, respectively. The resultant fit is shown in
Fig.~\ref{phot_1450}. When fitting a blackbody to the $IJHK$
magnitudes, a temperature of 2320~K has been obtained and, correspondingly,
an effective radius of $\sim 905\,R_\odot$ and a luminosity of $\sim 2.2
\times 10^4\,L_\odot$. The blackbody fit (with the contribution from the B3V
companion added) is shown as a dashed curve in
Fig.~\ref{phot_1450}. Thus the estimates of $T_{\rm eff}$ obtained
from the photometric measurements are well consistent with
the effective temperature we used to model the cold photosphere.

The photosphere model spectrum was then used
as an underlying spectrum absorbed in the outflowing molecular gas. In
radiative transfer modeling of absorption in the outflow (scattering
was not included), the gas was
treated as a plane-parallel homogeneous slab of matter. In this one-dimensional 
approach three parameters characterize the outflowing gas, i.e. 
column density of an absorbing molecule, excitation temperature
(the same for rotational and vibrational components), and a 
radial velocity of the slab (no velocity gradient was
implemented). Finally, to the resultant spectrum a photospheric 
contribution of a B3V star was added (a synthetic spectrum of 
T$_{\rm eff}$\,=\,18\,000\,K and $\log{g} = 4.0$ was used for this purpose
-- see Sect.~\ref{hotstar}).

It should be noted that the above approach was not meant to reproduce the observed
spectrum in the whole spectral range. Simulations were
performed separately for different narrow spectral regions and usually for one
molecule at once. It allowed us to
identify and study individual bands dominating in a given spectral range. 
With this modeling procedure rough estimates of physical parameters of
the outflowing material (excitation temperature, column densities, radial
velocity) were also possible in a number of cases.

A list of the identified molecular bands is presented in Table~\ref{TabMol}. The
first column of the table gives observed wavelengths of features. In most cases this is the
wavelength of a bandhead formed by the R branch, but in a few cases
this corresponds to a local minimum of the absorption feature  
formed by Q or P branches; otherwise an empty space is left. 
Column (2) contains laboratory wavelengths of the features. Columns
(3), (4), and (5) specify the molecule name, the electronic
system with a vibrational band identification, and the branch designation,
respectively. Column (6) gives a heliocentric velocity 
of the feature, if a clear bandhead is observed. 
References to appropriate molecular data and comments on individual
features can be found in the last two columns. Most of the identified
molecular features are also indicated on the spectrograms in
Figs.\ref{specB}--\ref{specR}. 

The spectrum is dominated by strong TiO absorption bands, which are
present in all parts of the HIRES spectrum. Prominent are also
bands of VO, AlO, ScO, and YO. In the following subsections, we describe
in some detail the identified features for each of the molecules.

\subsection{TiO}\label{tio}

Much effort was devoted to understand electronic transitions of TiO 
in laboratory \citep{PHI73}, in astronomical observations
\citep{RAM99}, as well as in theoretical computations
\citep{SCH98}. In the identification procedure,
we mainly consulted databases of line positions computed from
laboratory measurements \citep{JOR94,PLE98,SCH98}. These line
lists also provide data for bands not measured in laboratory and
in these cases discrepancies between different sources are often
significant. The lack of accurate spectroscopic data disables a detailed
analysis of some TiO bands, e.g. the $\gamma$' (3,0) band close to
5350\,\AA\ and higher overtones of $\gamma$' bands. 

We have identified three
electronic systems of TiO in the spectrum of V838  
Mon, i.e., $\alpha$ (C$^3\Delta-{\rm X}^3\Delta$), $\gamma$
(A$^3\Phi-{\rm X}^3\Delta$), and $\gamma$' (B$^3\Pi-{\rm X}^3\Delta$).
The molecular bands of the $\gamma$ and $\gamma$' systems have
generally a
characteristic triple structure formed by their main branches.
In the case of the $\alpha$ transitions, the separation of branches is 
so small that the three components overlap and the triple
structure is not seen. Additionally, in some bands of the
$\gamma$ and $\gamma$' systems, relatively strong satellite branches
are observed, e.g. the R$_{31}$ branch of the $\gamma$'
(1,0) band at 5827\,\AA. Finally, we have also identified
forbidden transitions of TiO in the spectrum. The latter
point is discussed in Sect.~\ref{forbid_TiO}.  

The observed features of TiO arise both in the stellar photosphere and in
the outflow. Although the latter component dominates formation of the
observed bands, it would be impossible to satisfactorily reproduce a number of
the observed features without a contribution from the
photosphere. Unfortunately, we have not been able to set valuable constraints
on the effective temperature, since there are only a few regions
in the spectrum, where the stellar component dominates. Performed
simulations show, however, that the selected atmospheric model with T$_{\rm
  eff}$\,=\,2400\,K works good in reproducing high excitation features seen
in the spectrum (see below).

Nonetheless, the most prominent absorptions of TiO seen in the
spectrum of V838 Mon arise in a low excitation environment.  
A comparison between the observations and spectra computed from 
our slab model with different excitation
temperatures suggests a mean value of T$_{\rm ex}$\,=\,500$\pm$100\,K.
Adopting this temperature 
and comparing the observed spectrum to the simulations of the
$\gamma'$ (2,0), (1,0), (0,0), and $\gamma$ (2,0) bands, an estimate of the
TiO column density of $\sim$16.8~dex~cm$^{-2}$ has been obtained. Assuming 
that all Ti is locked in TiO and taking a Ti atomic
abundance of 10$^{-7}$ relative to hydrogen, a column density of
hydrogen of $\sim$23.8~dex~cm$^{-2}$ has been derived.

A characteristic feature of the observed TiO spectrum is a strong
saturation of bands originating from the ground vibrational
state. The effect is clearly seen in the $\gamma$' (0,0) and (1,0) bands.
In the $\gamma$' (2,0) band the saturation is weaker and completely disappears 
in the (3,0) band. Similarly, a strong saturation is seen in the
$\alpha$ (0,0), (1,0), (2,0) bands, and probably also in the (3,0) and
(4,0) bands. In the red part of the analyzed spectrum, the $\gamma$
(0,0) and (1,0) bands are saturated. Due to the strong saturation 
in the $\gamma$ and $\gamma$' systems the triple structure of the
bands can be seen only through weak emission-like features that
appear between the saturated components, see e.g. the $\gamma$ (0,0) and
(1,0) bands at 6670\,\AA\ and 7100\,\AA\ (Fig.~\ref{specR}).

\begin{figure}[!bt]
\begin{center}   
\includegraphics[scale=0.65, trim= 20 0 0 0]{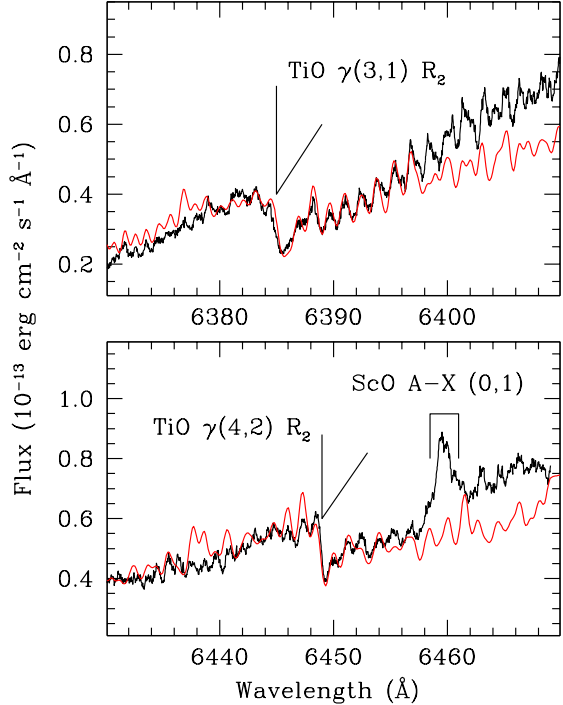}
\caption{Rotational structure of highly excited levels of two
  $\gamma$ bands of TiO. The observed spectrum (black line) is
  compared with simulations (red line). 
In order to fit the simulated rotational features to those
  in the observed spectrum, the simulated spectra have been shifted to
  a velocity of V$_h$\,=\,$58$\,km\,s$^{-1}$ and smoothed with a
  Gaussian with FWHM\,=\,34\,km\,s$^{-1}$. An emission
  feature of ScO can easily be seen in the bottom panel (ScO was
  not included in the simulations shown here).}
\label{fig_rot}
\end{center}   
\end{figure}

Radial velocities of some TiO bands can be determined by comparing
rotational features of the observed bands with simulated spectra. 
Although the rotational
structure is completely smoothed in the observed bandheads, 
it is recognizable in those parts of the spectrum, where only
absorptions from higher rotational levels are present. Two examples of such rotational
features are shown in Fig.~\ref{fig_rot}, where a simulated spectrum is overplotted
on the observed one. The cold slab had a excitation temperature of 500\,K in
the simulations. Note however that the contribution from the slab is
negligible in the $\gamma$(3,1) and (4,2) bands. To fit the rotational 
components the modelled spectrum has been shifted to
V$_h$\,=\,58$\pm$5\,km\,s$^{-1}$. Since these rotational features coming from high
excitation levels are predominantly formed in the
atmosphere of the central star, the derived
velocity is expected to be close to the radial velocity of the star. 

\begin{figure}[!bt]
\begin{center}
\includegraphics[scale=0.7, trim=20 0 0 40]{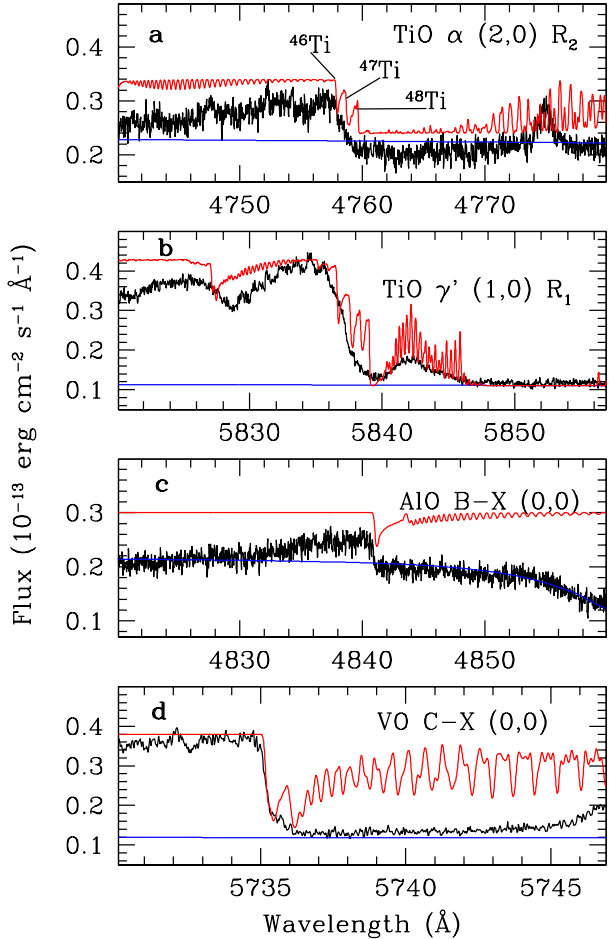}
\caption{Examples of observed (black line) bandheads of TiO, AlO, and
   VO, together with their simulated rotational profiles (red). On
   each panel a synthetic spectrum of a B3V star is also shown (blue). The rotational profiles 
   were shifted in velocity to fit the observed bandheads.   
~{\bf (a)}
  The $\alpha$ (2,0) bandhead of TiO. 
  The three
   heads clearly distinct in the simulated profile correspond to
   $^{46}$TiO, $^{47}$TiO, and $^{48}$TiO isotopomers. The
   profile was shifted in velocity by $-67\,$km\,s$^{-1}$.
~{\bf (b)}
  The $\gamma$' (1,0) bandhead of TiO. This
  head is formed by multiple strong satellite branches and defining a 
   reference laboratory position of the blue edge of the
   absorption feature is problematic. 
~{\bf (c)}
  The sharp B--X (1,0) bandhead of AlO. In this case the reference laboratory
   position is well defined.
  The velocity shift is $-82$\,km\,s$^{-1}$. 
~{\bf (d)} 
 The C--X (0,0) bandhead of VO. As seen in the computed profile, this
   head has a double structure when the band is not very
  saturated. The velocity shift is $-78$\,km\,s$^{-1}$.  
} 
\label{fig_wind}
\end{center}
\end{figure}

Radial velocities can also be determined from the observed positions of bandheads. In
the case of the TiO bands present in our spectrum formation of a bandhead  
is often strongly influenced by presence of satellite bands and
a non-negligible isotopic shift in wavelength. 
An example is
presented in panel (a) of Fig.~\ref{fig_wind}.  
This head, belonging to the $\alpha$ (2,0) band, is formed by the R$_2$
branch and a contribution from satellite bands is small in this
case. However, as can be seen in the rotational profile\footnote{A
  rotational profile means here a computed and smoothed profile of
  normalized absorption features of a band and it should not be
  confused with a full simulation from
  our model involving stellar photosphere and an absorbing cool slab
  as described in Sect.~\ref{molecules}.}
overplotted on the spectrum in Fig.\ref{fig_wind}, the head is
actually formed by multiple components. The strongest ones belong to the three main TiO
isotopomers, i.e., $^{46}$Ti$^{16}$O, 
$^{47}$Ti$^{16}$O, and $^{48}$Ti$^{16}$O. For solar Ti
isotopic ratios, i.e. 9.3,10.1,13.4,13.8 for 48/46,48/47,48/49,48/50,
respectively \citep{COW95}, all the isotopic components
should significantly contribute to the shape of the $\alpha$ (2,0)
bandhead. Since they are not distinguishable in the observed smooth
head, only a position of the most blueshifted component belonging to
the $^{46}$TiO isotopomer can be measured with a reasonable accuracy. 
In this way a velocity of V$_h=\,-67$\,km\,s$^{-1}$ has been inferred. 

Another example of a TiO complex bandhead is shown in panel (b) of
Fig.~\ref{fig_wind}. As seen in the rotational profile, the structure
of the $\gamma$' (1,0) bandhead is even more elaborate than in the
previous example. This is due to the presence of multiple heads formed
by satellite branches. The observed head is 
again very smooth and individual components, clearly distinguishable in the 
overplotted rotational profile, are not seen in the observed
spectrum. An estimate of the outflow velocity is very uncertain in
this case, but by fitting the rotational profile to the overall 
absorption profile of the band, a velocity of
V$_h = -58 \pm 10$\,km\,s$^{-1}$ has been derived.

 As can be found in Table~\ref{TabMol}, the most blueshifted
molecular component is observed in the $\alpha$ (0,0) band at
5164.6\,\AA. A velocity of $-125$\,km\,s$^{-1}$ has been derived
assuming that the blue 
edge of the head is formed by the $^{46}$TiO isotopomer, instead of the
most abundant $^{48}$TiO. A similar velocity was found for the
$\alpha$ (1,0) band at 4952.6\,\AA. 

The derived radial velocities of TiO features are in a range
between $58\,$km\,s$^{-1}$ and $-125\,$km\,s$^{-1}$. It is reasonable
to assume that the molecular absorption arises in an outflow with
a velocity gradient. This can explain different velocities
of the bands with different excitation requirements, and can
also account for the observed smooth structure of the bandheads.

\subsection{AlO}

We based our identification procedure of the AlO bands on an analysis of
the B$^{2}\Sigma^{+}$\,--\,X$^{2}\Sigma^{+}$ system performed in \cite{COX85}. 
Absolute strengths of the electronic transitions ware taken from \cite{PAR83}.

The blue-green (B$^{2}\Sigma^{+}$\,--\,X$^{2}\Sigma^{+}$) band system of AlO
is evidently present in the analyzed spectrum. The most prominent
feature is the saturated (0,0) band with a main head at 4842\,\AA, see
panel (c) in Fig.~\ref{fig_wind}. Three bandheads of different
   branches are close enough to form this relatively sharp head. In
   this case a radial velocity can be measured accurately and we have found
   $V_h = -82 \pm 6$~km~s$^{-1}$. 
  
Higher vibrational overtones, $\Delta$v\,=\,+1 and $\Delta$v\,=\,+2, of the
B--X system can also be recognized in the spectrum of V838 Mon. 
The more intense $\Delta$v\,=\,+1 overtone is seen up to the (6,5)
band around 4754\,\AA. The radial velocity of the bandheads of the first overtone
cannot be defined accurately due to the poor quality of the spectrum
in the blue range. However, the wavelengths of bands from higher vibrational
levels imply a velocity of $V_h\simeq 58$\,km\,s$^{-1}$. The
$\Delta$v\,=\,+2 overtone is only hardly seen in the spectrum. The most
intense (2,0) band is contaminated by the \ion{He}{1} 4471\,\AA\
line of the B3V star.  

\begin{figure}[!bt]
\begin{center}   
\includegraphics[scale=0.42, trim=20 0 0 0]{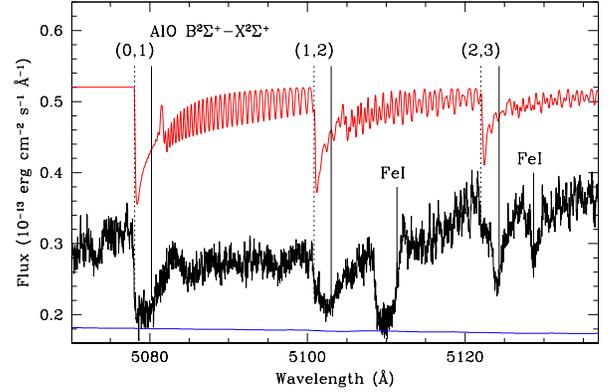} 
\caption{Observed spectrum (black line) of the A--X $\Delta$v=$-$1
  subtone of AlO together with its rotational profile computed for
  an excitation temperature of 1500\,K (red). The profile was shifted in
  velocity by $-82$\,km\,$^{-1}$. The dotted lines mark positions of 
  the bandheads if shifted to $V_h = -82$\,km\,s$^{-1}$, while the solid
  vertical lines are drawn for $V_h = 58$\,km\,s$^{-1}$. The blue line marks the
  flux level of the B3V companion.}
\label{fig_AlO}
\end{center}
\end{figure}

In the green part of the spectrum, vibrational bands belonging to the
$\Delta$v\,=\,$-1$ sequence are clearly visible. The bands are shown in
Fig.~\ref{fig_AlO}.  
The (0,1) band at 5079\,\AA\ is particularly strong. Its bandhead
is saturated and shifted to $V_h = -82$\,km\,s$^{-1}$. One can also
distinguish other members of the same overtone originating from higher
vibrational levels up to v"=\,4. According to our
simulations, these bands, to be observed, require a rather high
excitation temperature of about 1500\,K, which suggests that they
originate in, or close to, the atmosphere of the star.
However, sharp bandheads belonging to transitions with
v" $\leq$\,3  are shifted to a velocity $V_h = -82$\,km\,s$^{-1}$,
which is inconsistent with the stellar velocity. The velocity rather
indicates that the absorbing gas is located in outer parts
of the outflow. Note that this velocity is almost the same as 
that of the narrow absorbing component seen in several atomic lines, 
e.g. \ion{Rb}{1} (see Sect.~\ref{pcyg}).    

\subsection{VO}

The spectrum of V838 Mon shows strong vibrational bands of 
the C$^{4}\Sigma^{-} -$\,X$^{4}\Sigma^{-}$ and B$^{4}\Pi
-$\,X$^{4}\Sigma^{-}$ systems of VO. 
For the purpose of our simulations a line list for these
systems was computed using spectroscopic constants from \cite{CHE82}.
Rotational strength factors were taken from \cite{KOV} with corrections
applied after \cite{WHI}. In the case of the B--X (0,0), (1,0), and
(2,0) bands, a list of energy levels was generated by 
diagonalization of the perturbed Hamiltonian from \cite{ABB95} and \cite{CHE94}. 
The transformation matrices were next used to compute 
rotational strengths of individual lines to properly include effects
of perturbations. 
The rotational
structure of the (2,0) band was not analyzed in laboratory. Therefore
spectroscopic parameters for this band were extrapolated from the lower overtones.
Electronic strengths were taken from \cite{kll97}. Franck-Condon
factors were computed with the LeRoy's codes\footnote{\tt
  http://leroy.uwaterloo.ca/programs/} \citep{ler1,ler2}. 

As can be seen in panel (d) of Fig.~\ref{fig_wind}, the unsaturated C--X (0,0)
band has two heads. One is formed by the R$_{1}$ and R$_{4}$ 
branches, while the other, redshifted by 1\,\AA, is  formed by R$_{2}$ and R$_{3}$. 
Both are present in the observed spectrum and the former one appears at
a velocity of $-77$\,km\,s$^{-1}$. A regular ladder of rotational lines of the (0,0)
band formed by main P branches can be recognized starting at
5753\,\AA\ (see Fig.~\ref{specG}). Fortunately, precise laboratory wavelengths
are available for these lines and a detailed modeling of the band
was possible. The simulations suggest that this part of the spectrum
is formed at a low excitation temperature of $\sim$300\,K. The derived
velocity of the absorbing gas in the rotational transitions of the
(0,0) band is $V_h$\,=\,43\,km\,s$^{-1}$. From the heads of
higher transitions of the C--X system, namely the (1,0) and (2,0) bands,
velocities of $-77$\,km\,s$^{-1}$ and
$-45$\,km\,s$^{-1}$ were derived, respectively. 

\begin{figure}[!bt]
\begin{center}
\includegraphics[scale=0.42,trim=20 0 0 0]{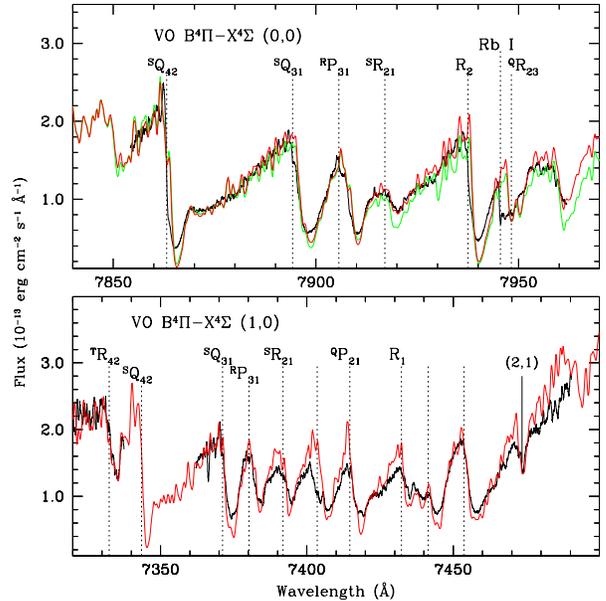}
\caption{Comparison between the observed (black line) and simulated
  spectra of the B--X bands of VO. 
{\bf Top:} 
  the (0,0) band
  was simulated with the excitation temperatures of 350\,K (red line)
  and 500\,K (green line). The simulated spectra were shifted
  to a velocity of --2\,km\,s$^{-1}$ and 
  smoothed with a Gaussian with  FWHM\,=\,65 km\,s$^{-1}$. The vertical lines mark
  positions of the bandheads for a radial velocity of --80\,km\,s$^{-1}$,
  which fits to the observed bandheads of $^S$Q$_{42}$ and
  R$_2$, and to the blue edge of the \ion{Rb}{1} profile. 
{\bf Bottom:} 
  the simulation of the (1,0) band obtained with two layers of the
  absorbing molecular gas with different temperatures, i.e. 500\,K and
  1500\,K. The colder layer was shifted in velocity to
  --2\,km\,s$^{-1}$, while the warmer one, treated in our radiative
  transfer modeling as a layer between the stellar atmosphere and the cold
  slab, was shifted to a velocity of
  58\,km\,s$^{-1}$. The dotted lines mark positions of the bandheads for a
  radial  velocity of --60\,km\,s$^{-1}$, which fits to the
  $^T$R$_{42}$ and $^S$Q$_{31}$ bandheads. The absorption feature
  at 7474\,\AA\ belongs to the (2,1) band.
}   
\label{VO_fig00}
\end{center}
\end{figure}

Additionally, within the B--X system, the bands (0,0), (1,0), and (2,0) were
identified.
The (0,0) and (1,0) bands,
together with their simulated spectra, are presented in
Fig.~\ref{VO_fig00}. Most of the features of these bands are only slightly
shifted from their laboratory wavelengths and the measured velocity is
$V_h$\,=\,$-2$\,km\,s$^{-1}$. A rather uncertain value of $\sim$350\,K was
deduced from the simulations as an excitation temperature of the
absorbing gas (mainly from the intensities of absorption features at
7910\,\AA\ and 7920\,\AA). Furthermore, in order to reproduce some features seen in the
spectra of the bands with higher vibrational numbers, e.g. the absorption feature
at 7474\,\AA\ belonging to the (2,1) band, it was necessary to add
an additional layer of absorbing VO gas with a rather high excitation
temperature, namely
T$_{\rm ex}$=1500\,K. This hotter component, treated in the simulations as a
slab between the stellar photosphere and the cold slab, 
appears at a velocity of $\sim$58\,km\,s$^{-1}$.

Our simulations of the C--X (0,0), B--X (0,0) and (1,0) bands give
satisfactory fits to the observations when a column density of the VO
absorber is of $\sim$16.4~dex~cm$^{-2}$. If we assume that all
V is locked in VO molecules and taking the solar abundance of
V relative to H of 10$^{-8}$, this corresponds to a column
density of H of $\sim$24.4~dex~cm$^{-2}$. The result is consistent with a
similar estimate performed from the TiO bands in Sect.~\ref{tio}.

\subsection{ScO}

The only electronic system of ScO present in the spectrum is the
A$^{2}\Pi - {\rm X}^{2}\Sigma^{+}$ system. Absorption features of the
(0,0), (1,1), (2,1) bands are clearly seen in the green part of the
spectrum. Moreover, one band of ScO,
i.e. the (0,1) band, was found in emission. For the purpose of
our simulations, line positions of the A--X (0,0) 
band were computed from the Hamiltonian diagonalization
parameterized by \cite{SAF72} and \cite{ADA68}. Higher vibrational levels
were approximated from data in NIST\footnote{\tt http://physics.nist.gov/PhysRefDat/ASD/
}. For the positions of transitions of the (1,0) band, \cite{RF86} was
consulted. Strengths of rotational lines were found in the approach
proposed in \cite{herbig74}. It should be noted, that although the line positions
of the (0,0) band are exact, their strengths are only approximate.

\begin{figure}[!bt]
\begin{center}   
\includegraphics[scale=0.4,trim=20 0 0 0]{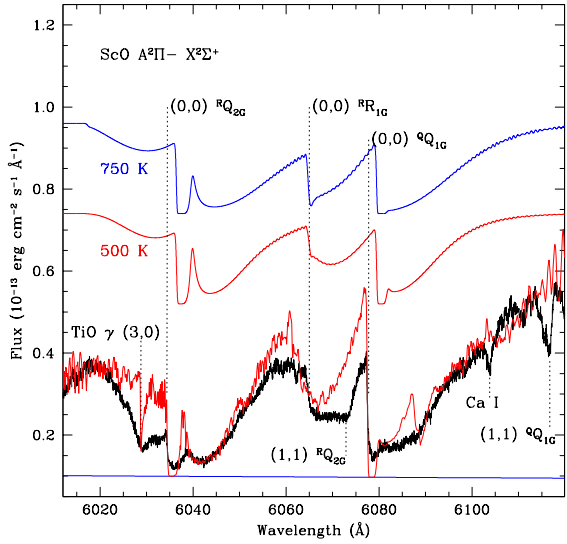}
\caption{Comparison between the observed and simulated spectrum of
  the A--X (1,0) band of ScO. The two upper
  plots are rotational profiles of the band, computed for
  excitation temperatures of 750\,K and 500\,K (blue and red,
  respectively). They are shifted to a velocity of
  58\,km\,s$^{-1}$. The observed
  spectrum is drawn with a black line. Overlaid on it in red is a simulated spectrum
  with a molecular slab absorbing the light of the cool star as described in
  Sect.~\ref{molecules}. The absorbing layer
  has an excitation temperature of 500\,K and is shifted in velocity
  by --55\,km\,s$^{-1}$. The simulated spectrum is smoothed with a Gaussian with
  FWHM\,=\,32\,km\,s$^{-1}$. The flux level of the B3V star is indicated 
  with the blue line.}  
\label{fig_ScO}
\end{center}
\end{figure}

Figure~\ref{fig_ScO} shows a region of the observed spectrum covering two
subbands of the (0,0) band. Simulations of this spectral region
suggest the presence of at least two different velocity
components. The bandhead formed by the $^2\Pi_{3/2}$ $^RQ_{2G} + 
{^R}R_{2G}$ (at 6036\,\AA) and $^2\Pi_{1/2}$ ${^Q}Q_{1G} + {^Q}R_{1G}$
(at 6079\,\AA) branches are shifted to $-$55\,km\,s$^{-1}$ (as measured for the
lowest part of the profile). The second velocity component at
45\,km\,s$^{-1}$ was found
from the position of the head formed by the $^2\Pi_{1/2}$ $^RR_{1G}$
branch at 6064.17\,\AA. This latter head is formed at a rather high 
rotational number, i.e., N=58, contrary to the heads observed at
$-$55\,km\,s$^{-1}$, which are formed at N=28 and N=17. The relative
velocity shift can be understood as a result of decreasing population
of high rotational levels with increasing velocity in the observed
outflow. A lack of accurate laboratory wavelengths disables an
analysis of other absorption bands of ScO.  

Remarkable is the presence of the
A--X (0,1) $^2\Pi_{3/2}$\,$^{R}$Q$_{2G}$ and $^2\Pi_{1/2}\,^Q$Q$_{1G}$
subbands of ScO in emission at 6408\,\AA\ and 6458\,\AA, respectively
(see bottom panel of Fig.~\ref{fig_rot} for the latter subband).  
The observed velocity of this features agrees well with the
velocity of emission components seen in the atomic P-Cyg profiles
(see Sect.~\ref{pcyg}). The origin of this molecular emission can be
explained by fluorescent pumping of the upper electronic level by
absorption in the (0,0) 6090\,\AA\ and 6132\,\AA\ subbands, clearly seen in
the spectrum of V838 Mon. 

\subsection{YO}

All the features of YO identified in the spectrum of V838 Mon
belong to the A$^2\Pi$--X$^2\Sigma^{+}$ system. A list of the
transitions used in our identification procedure was prepared on the
basis of data provided in \cite{BER83}. There is another well known
system of YO, i.e. B$^2\Pi-{\rm X}^2\Sigma^{+}$ around
4819\,\AA\ \citep{BBL79}, but it falls into a spectral region 
of saturated TiO bands and was not found in our spectrum. 

\begin{figure}[!bt]
\begin{center}
\includegraphics[scale=0.4]{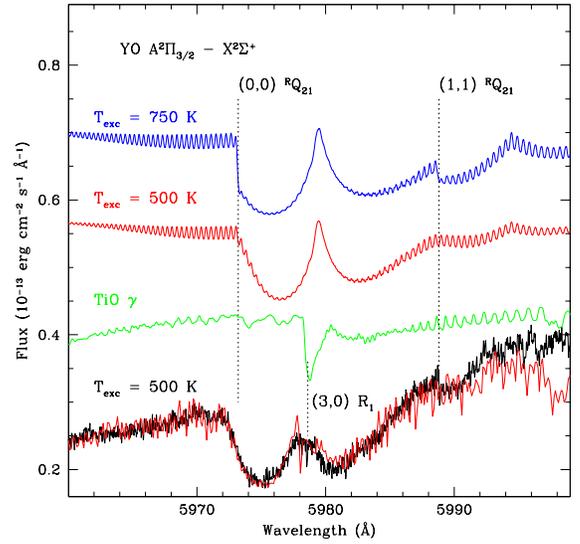}
\caption{Same as in Fig.~\ref{fig_ScO} but for the fine components of
  the electronic transitions of the A$^{2}\Pi_{3/2}-{\rm X}^{2}\Sigma^{+}$
  feature of YO. The final simulation (bottom) includes an extra absorption of
  TiO, which is shown with a green line. The simulation
  was performed with a slab of 500\,K and
  the stellar photosphere. All the computed spectra
  were shifted to a velocity of 58\,km\,s$^{-1}$ and smoothed with a
  Gaussian with FWHM\,=\,37\,km\,s$^{-1}$.} 
\label{fig_YO}
\end{center}
\end{figure}

We performed a detailed simulation of the A$^2\Pi_{3/2}-{\rm
  X}^2\Sigma^{+}$ transition. The results of this simulations are shown
  in Fig.~\ref{fig_YO}. The upper two plots in Fig.~\ref{fig_YO} are
  rotational profiles of the transition obtained for two excitation
  temperatures, 750\,K and 500\,K. It can be seen that a sharp edge of
  the (0,0) bandhead at 5973\,\AA\ is formed with the higher temperature. In
  the observed spectrum this head is very smooth and resembles more
  the feature seen in the profile for 500\,K, and this is a value we
  took for further simulations. It should be 
  noted, however, that the observed profile of the band can be
  influenced by the physical structure of the wind, which is ignored in our
simulations; in particular, a velocity gradient in the molecular
  outflow can smooth the shape of the bandheads, and even if the
  temperature is high, no sharp head would be observed. Nonetheless,
  to perform full simulations of the  A$^2\Pi_{3/2}-{\rm
  X}^2\Sigma^{+}$  transition of YO it was necessary to include the $\gamma$
  (3,0) band of TiO as an additional important absorber in this
  spectral range. The green line in
  Fig.~\ref{fig_YO} shows the TiO contribution to the
  observed spectrum. 
  The final simulation which includes absorption of YO
  and TiO with an excitation temperature of 500\,K is shown at the bottom of
  Fig.~\ref{fig_YO}. 
From the overall fit of
  the (0,0) band to the full simulation a velocity of
  $V_h$\,=\,--8\,km\,s$^{-1}$ has been derived.

Note however that a sharp (1,1) bandhead formed by the $^RQ_{21}$ branch at
5988\,\AA\ appears at higher excitation temperatures, at
least 750\,K, as can be seen from 
Fig.~\ref{fig_YO}. Such a sharp feature seems to be present in the
observed spectrum, suggesting that this absorption feature is formed in
gas with a somewhat higher temperature than the one found for the
(0,0) bandhead. This 
is confirmed by the measured velocity of the (1,1) bandhead,
$V_h$\,=\,58\,km\,s$^{-1}$, which is the same as that of other
high-excitation features found in the spectrum. 

\subsection{TiO -- forbidden transitions}\label{forbid_TiO}

\begin{figure}[!bt]
\begin{center}
\includegraphics[scale=0.45,trim=20 0 0 0]{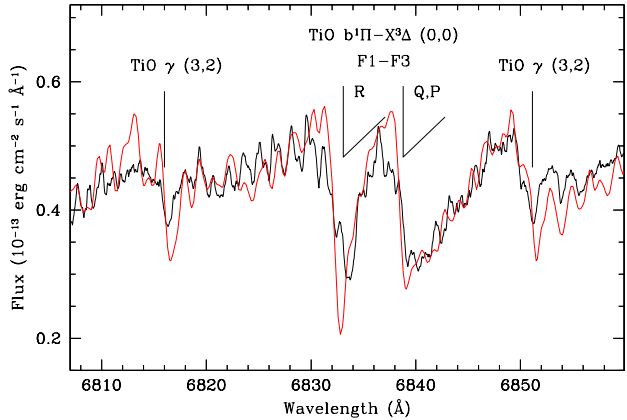}
\caption{Observed and simulated spectrum of the forbidden band
  b$^{1}\Pi$--X$^{3}\Delta$ (0,0) of TiO. The simulation (red line) 
  was obtained with an excitation
  temperature of 500\,K. It was shifted to a velocity
  $V_h$\,=\,--82\,km\,s$^{-1}$ and smoothed with a Gaussian with
  FWHM\,=\,32\,km\,s$^{-1}$.} 
\label{TiO_f_fig}
\end{center}
\end{figure}

Very interesting from the spectroscopical point of view is the presence of 
forbidden transitions of TiO, i.e. the c$^{1}\Phi$--X$^{3}\Delta$
(0,0) band around 4737\,\AA\ and the b$^{1}\Pi$--X$^{3}\Delta$ (0,0)
band at 6832\,\AA. The stronger b--X band is shown in detail in
Fig.~\ref{TiO_f_fig}. The intensities of the bands were predicted in
{\it ab initio} calculations of ro-vibrational energy levels of the 13
lowest electronic states of TiO by \cite{SCH98}. As shown in
Fig.~\ref{TiO_f_fig}, our simulation can reproduce the forbidden bands
quite well. The best fit indicates a velocity of
$V_h=-$80\,km\,s$^{-1}$ and an excitation temperature 
of about 500\,K. However, when the
column density found for other TiO bands (see  Sect.~\ref{tio}) is
assumed, oscillator strengths of the b$^{1}\Pi-$X$^{3}\Delta$ 
band must be enhanced by a factor of 70$\pm$35 in order to reproduce 
the observations. To our knowledge this is the
first, in astrophysics as well as in laboratory, observation of TiO
forbidden bands. 

\section{Photospheric spectrum of the B3V companion}\label{hotstar}
A number of broad absorption lines of the Balmer series, 
from H$\beta$ down to at least H{\small 12}, as well as a few \ion{He}{1}
lines, can clearly be seen in the blue part of our spectrum
(Fig.~\ref{specB}). These features are attributed to the B3V companion.
We compared the observed spectrum to a grid of
synthetic spectra {\small BSTAR2006}, generated with the {\it
  Synspec} program \citep{synspec}. 
Solar metallicity models, with effective temperatures between $15\,000$
and $30\,000$~K 
and gravities in the range $\log{g} = 2.0 - 4.75$ 
were considered. A turbulent velocity of 2~km~s$^{-1}$ was adopted. 
Good fits to the
observed spectrum were obtained for T$_{\rm eff} = 18\,000 \pm 2\,000$~K and
$\log{g} = 4\pm0.75$. 
The low signal-to-noise ratio in the part dominated by the B3V star
and uncertainties in the flux calibration did not allow us to obtain more
precise estimates of the star parameters. In any case, the above values agree
well with those expected for a B3V star.
In the following discussion we use the synthetic spectrum with
T$_{\rm eff}=18\,000$~K and $\log{g} = 4.0$. This spectrum, broadened with a rotational
velocity of 250~km~s$^{-1}$ (see below) and shifted to a radial velocity
found for the hot star (see below), is overplotted on the observed spectrogram of
V838~Mon in Figs.~\ref{specB}--\ref{specR}. 

A comparison of the observed spectrum to the synthetic one
shows that the absorption lines
of the B3V companion are rotationally broadened.
In order to find the rotational velocity, we
convolved the synthetic profiles with a Gaussian profile
broadened with a grid of
rotational velocities (we made use of the {\it rotin3} program\footnote{\tt
  http://nova.astro.umd.edu/Synspec43/ synspec-frames-rotin.html}). 
Note that this procedure does not take into account limb darkening nor
gravitational darkening, which can be important for profile formation 
in a rapidly rotating star. Lines 
usually used to investigate rotational broadening of B-type stars, i.e. 
\ion{He}{1} 4471~\AA\ and \ion{Mg}{2} 4481~\AA, are weak, noisy,
and affected by 
emission lines and/or molecular bands in our spectrum. However, we have
found that in the green part of the spectrum the \ion{He}{1} 5876~\AA\ line 
is relatively strong and has a ``clean'' profile. This green spectral region
is dominated by the cool photosphere but, fortunately, in the spectral vicinity
of the $\lambda$5876 line the flux from this component is practically reduced
to zero due to a saturated TiO band (see Fig.~\ref{specG}) and the line
from the B3V component is well seen.
Thus, when determining the rotational velocity of the B3V component we
relied mainly (but not only) on the profile of \ion{He}{1} 5876~\AA.
Using the T$_{\rm eff}$=$18\,000$~K, $\log{g}$=4.0 template we found
a projected rotational velocity $V \sin{i} =
250\pm50$~km~s$^{-1}$. This is a rather large
value\footnote{For instance, in \cite{abt} one finds
               that the mean projected rotational velocity for 106
	       dwarfs with spectral types B3$-$B5 is $108\pm8$~km~s$^{-1}$,
	       and only 2.4\% of the stars
	       have $V \sin{i} \geq 300$~km~s$^{-1}$.} 
and indicates that the rotation axis is close to the plane of the sky.

We also attempted to measure the radial velocity of the hot
star. We used the Fourier cross-correlation method \citep{ccf} as it
is implemented in the IRAF's 
task {\it rvcor}. As a template, we used our synthetic spectrum broadened
to the rotation velocity of 250~km~s$^{-1}$. The cross-correlation was
performed for the region between 3815~\AA\ and 4180~\AA, with
all emissions and ISM absorptions in this range removed manually. (We
performed tests with ranges including H$\beta$ and H$\gamma$, but they did not
give reliable results. These spectral regions are too strongly contaminated by atomic
emissions and molecular absorptions. Also, including the very noisy
shortward part of the spectrum spoils the cross-correlation function
and gives unrealistic results.)
As a result we obtained a heliocentric velocity
of $37.0\pm16.8$~km~s$^{-1}$ [with a Tonry \& Davis' (1979)
$r$ value of 18.0],
or correspondingly $19.6\pm 16.8$~km~s$^{-1}$
in to the local standard of rest. 
The error value was taken directly from the output of 
the {\it rvcor} procedure and, in our opinion, it is slightly overestimated.

\section{Interstellar features}\label{ism}

\begin{figure}
  \centering
  \includegraphics[scale=.7]{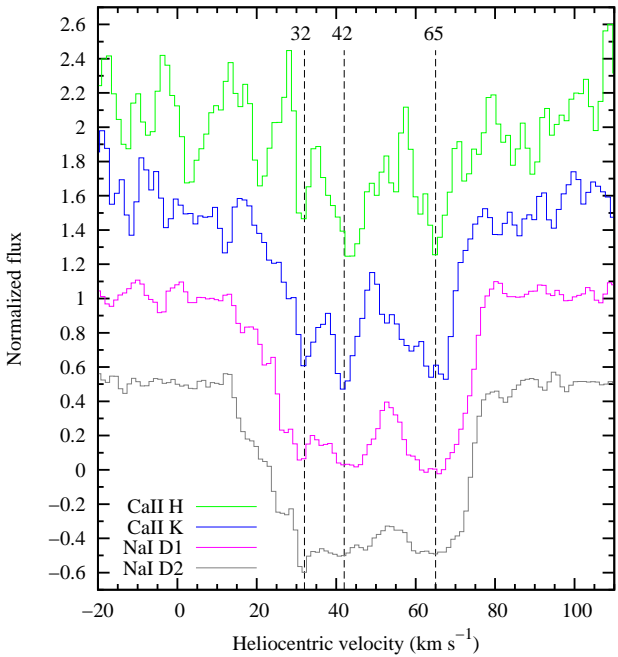}
  \caption{Interstellar doublets of \ion{Ca}{2} and \ion{Na}{1}. The
    curves, except \ion{Na}{1}~D$_1$, are shifted in the vertical scale 
    to better show the profile structures. 
    The \ion{Ca}{2}~H profile is noisy as the line is
    observed near the core of the H$\epsilon$ absorption, where the stellar flux is
    very low (see Fig.~\ref{specB}).
    Three main absorption
  components at 32, 42, and 65~km~s$^{-1}$ are marked with dashed
  vertical lines.}
\label{ism_fig}
\end{figure}

Several interstellar absorption features can be identified in our spectrum.
The strongest ones are those of \ion{Ca}{2} H, K and \ion{Na}{1} D$_1$,
D$_2$. They are shown in Fig.~\ref{ism_fig}. Interstellar components can also be
found in the deep P-Cyg absorption of \ion{K}{1} at 7699~\AA\, (see
upper panel in Fig.~\ref{pcyg_fig}). The \ion{Na}{1} and \ion{K}{1}
interstellar absorptions in the spectrum of V838~Mon have already been studied 
in a number of papers \citep[e.g.][]{kipp04,mun05}.
As can be seen from Fig.~\ref{ism_fig}, the profiles of the \ion{Ca}{2} lines, 
not described in the literature so far, have a similar structure to those of
\ion{Na}{1}. All the four profiles displayed in Fig.~\ref{ism_fig} show three
components at 32, 42, and 65~km~s$^{-1}$, similarly as observed in
\cite{mun05}. We note, however, that unlike previous studies, where 
the interstellar lines were observed in the spectrum of V838~Mon itself, the
features presented in Fig.~\ref{ism_fig} are observed in the spectrum of the
B3V companion. As can be seen from Figs.~\ref{specB} and \ref{specG} the
local continuum near the discussed lines comes from the B3V companion. At
the wavelengths of the \ion{Ca}{2} lines V838~Mon is too cool to emit any
significant radiation, while at the wavelengths of the \ion{Na}{1} lines an
absorption band of TiO is so strong that it eats up practically all the flux
from the cool supergiant. The fact that the interstellar line profiles in
the spectrum of the B3V companion are very similar to those in the spectrum
of V838~Mon itself is consistent with the idea that both stars are related,
at least that they are at a similar distance.

We have identified six diffuse interstellar bands (DIBs) in our spectrum,
i.e., at
$\lambda\lambda$4428, 5780, 5797, 6284, 6614, and 7224 [the
central wavelengths are taken from \citet{herbig}]. They all are redshifted
to a heliocentric velocity of $\sim$60~km~s$^{-1}$. It is well known that
equivalent widths of the features at 5780 and 5797~\AA\
correlate very well with interstellar extinction. Using conversion formulae
from \cite{herbig75} we derived $E_{B-V}$ of 0.95 and 0.90
for the DIBs at 5780~\AA\ and 5797~\AA, respectively. Taking
the relation from \cite{herbig93}, which is based on a
larger number of data points, gives $E_{B-V} \simeq 1.00$ for both 
features. Given considerable uncertainty of these estimates, due to
uncertainties in the equivalent width measurements and the intrinsic scatter 
in the relations used, our result is
consistent with $E_{B-V} \simeq 0.9$ obtained in other studies 
[as summarized in \citet{tyl05}].

\section{Summary}
We have obtained a spectrum of V838~Mon on 2005 October 13 using the
Keck/HIRES instrument. The spectrum of a resolving power of $\sim 34\,000$
covered a wavelength range of 3720~--~7960~\AA. The spectrum is very complex
and shows numerous spectroscopic features which can be identified as arising
from different components or environments. V838~Mon itself, i.e. the object which 
erupted in 2002, is seen mainly in the green and red parts of the spectrum 
as a cool continuum with strong, broad and complex
molecular absorptions. Numerous atomic lines showing P-Cyg profiles or
broad absorption features are also present in this component.
The blue part of the spectrum is dominated by a
photospheric spectrum of the B3V companion, showing broad absorptions lines
of \ion{H}{1} and \ion{He}{1}. In the blue and green parts of the spectrum,
numerous emission lines can be identified. These mostly arise from forbidden
transitions in singly ionized iron. Finally, several absorption features of
interstellar origin have been identified. These include atomic resonance
lines as well as DIBs.

The numerous [\ion{Fe}{2}] emission lines observed in our spectrum evidently arise
from rarefied and (partly) ionized matter. All the lines
show the same profile, which can be well fitted with a Lorentzian profile.
The lines are centred at a heliocentric radial velocity of
$\sim$13~km\,s$^{-1}$ and have a typical FWHM of $\sim$80~km\,s$^{-1}$.

In the cores of the photospheric absorption lines of the B3V companion we
see weak emission features. The double feature in the H$\beta$ line can be
interpreted as a single emission with a narrow absorption superimposed on
it. The derived parameters of the emission component are close to those of the
[\ion{Fe}{2}] lines. This suggests that the Balmer emission features arise
from the same environment as the [\ion{Fe}{2}] emission lines.

The spectrum displays a number of strong atomic, mostly resonance, lines with 
P-Cyg profiles. Evidently V838~Mon, being almost 4 years after its
outburst, continues losing mass.
The emission component typically peaks at a heliocentric
velocity of $\sim$80~km\,s$^{-1}$, while the absorption component usually
extends from $\sim$50~km\,s$^{-1}$ to $\sim -100$~km\,s$^{-1}$. The
absorption component often shows a complex profile. In particular, in a few lines
a strong and narrow absorption component is clearly seen at a velocity of 
$\sim -80$~km\,s$^{-1}$.
A few atomic lines were observed to show pure absorption profiles.

The spectrum from the cool component is dominated by numerous, often very
strong, molecular bands in absorption. Some bands are so deep that the flux
from the cool component practically goes to zero. In these cases we have a
sort of spectral windows, through which we can see the B3V companion, 
even in the red part of the spectrum. All the identified bands are from
oxides. They include TiO, VO, AlO, ScO, and YO. 

Our analysis shows that the bands are
formed in different regions. Some of them, especially those from highly
excited states, are formed, at least partially, in the photospheric regions.
The temperature in these layers is as high as $\sim$2400~K. These bands show a
heliocentric radial velocity of $\sim$58~km\,s$^{-1}$, which we identify as
a radial velocity of V838~Mon itself.

Most of the bands are formed in regions significantly cooler than the
photospheric ones. The vibrational temperatures derived from some bands go
down to $\sim$500~K. These are usually very strong bands arising from ground
states. The radial velocity of these bands is usually much lower than that
of the photospheric regions and goes down to $\sim -125$~km\,s$^{-1}$. Overall, our analysis of the molecular bands shows that they are mostly
formed in outflowing matter with a significant velocity gradient.

We have detected bands arising from forbidden transitions in TiO. As far as
we know, these are first observations of TiO forbidden transitions in
astrophysical objects.

The observed spectrum of the B-type companion, seen primarily in the blue
part of our spectrum, can be fitted with a model atmosphere spectrum with
solar abundances, effective temperature of $\sim$18\,000~K, and a gravity,
log\,$g \simeq 4.0$. These values agree well with those expected for a B3V
star. The absorption lines are significantly broadened, which we interpret
as a rotational broadening with a velocity, $V\,{\rm sin}\,i \simeq
250$~km\,s$^{-1}$. We have derived a heliocentric radial velocity of the B3V
companion of $37\pm17$~km\,s$^{-1}$.

Several interstellar absorption features are seen in our spectrum. They
include lines of \ion{Ca}{2}, \ion{Na}{1}, and \ion{K}{1}, as well as
DIBs. The lines of \ion{Ca}{2} and \ion{Na}{1} are
observed in the spectrum of the B3V companion. Their profiles are very
similar to those observed in the spectrum of V838~Mon during the 2002
outburst. This result is consistent with the idea that V838~Mon and the B3V
companion are related objects. From equivalent widths of diffuse interstellar 
bands we have estimated an interstellar reddening of $E_{B-V} \simeq 0.9 -
1.0$, which agrees well with interstellar extinction values obtained by
other authors using other methods.

A detailed analysis and interpretation of the data presented in the present
paper, as well as of data from other sources, will be done in a forthcoming
paper.

\begin{acknowledgements} 
MS, RT, and TK were
supported by the Polish Ministry of Science and Higher Education under
grant no. N203 004 32/0448, for which they are grateful.
TK wishes to thank T. Tomov for his helpful advises on data reduction and
spectrum analysis with IRAF. MK was supported by the Foundation for Polish Science through a FOCUS
grant and fellowship, by the Polish Ministry of 
Science and Higher Education through grants N203~005~32/0449 and 
1P03D~021~29. MG acknowledges support by Polish KBN
grant no. 1P03d~017~27 and SALT International Network grant
no. 76/E-60/SPB/MSN/P-03/DWM~35/2005-2007.  
This paper uses observations made at the South African Astronomical
Observatory (SAAO). \\
The authors wish to recognize and acknowledge the very significant
cultural role and reverence that the summit of Mauna Kea has always
had within the indigenous Hawaiian community.  We are most fortunate
to have the opportunity to conduct observations from this mountain.     
\end{acknowledgements}

{\it Facilities:} \facility{Keck:I (HIRES)}, \facility{Radcliffe (Grating Spectrograph)}



\begin{deluxetable*}{c c l c c}

\tablewidth{0pt}
\tabletypesize{\small}
\tablecaption{Identified emission lines\label{atom_em_tab}}
\tablehead{
\colhead{$\lambda_{\rm lab}^{\rm air}$ (\AA)}& 
\colhead{Ion}&
\colhead{Multiplet}&
\colhead{Flux}&
\colhead{Notes}}
\startdata
3883.820 &~[\ion{Fe}{2}]? & 24F &\nodata&a\\
3993.059 &~[\ion{Ni}{2}]? & 4F  &\nodata&\nodata\\
4101.734 & H$\delta$& 1   &\nodata&\nodata\\
4177.196 & [\ion{Fe}{2}]  & 21F & 3.355e--15 &b\\
4201.172 &~[\ion{Ni}{2}]? & 3F  &\nodata&\nodata\\
4243.969 & [\ion{Fe}{2}]  & 21F & 5.028e--14 &c,d\\
4244.813 & [\ion{Fe}{2}]  & 21F & 1.448e--14 &c,d\\
4276.829 & [\ion{Fe}{2}]  & 21F & 5.469e--14 &\nodata\\
4287.394 & [\ion{Fe}{2}]  & 7F  & 1.423e--13 &\nodata\\
4305.890 & [\ion{Fe}{2}]  & 21F & 1.292e--14 &\nodata\\
4319.619 & [\ion{Fe}{2}]  & 21F & 2.852e--14 &\nodata\\
4326.237 &~[\ion{Ni}{2}]? & 3F  &\nodata&\nodata\\
4340.464 & H$\gamma$& 1   &\nodata&\nodata\\
4346.852 & [\ion{Fe}{2}]  & 21F &\nodata&c\\
4352.778 & [\ion{Fe}{2}]  & 21F & 2.448e--14 &\nodata\\
4358.360 & [\ion{Fe}{2}]  & 21F & 2.842e--14 &d\\
4359.333 & [\ion{Fe}{2}]  & 7F  & 1.022e--13 &d\\
4372.427 & [\ion{Fe}{2}]  & 21F & 2.940e--14 &\nodata\\
4382.742 & [\ion{Fe}{2}]  & 6F  &\nodata&\nodata\\
4413.781 & [\ion{Fe}{2}]  & 7F  & 1.109e--13 &c,d\\
4416.266 & [\ion{Fe}{2}]  & 6F  & 5.579e--14 &c,d\\
4432.447 & [\ion{Fe}{2}]  & 6F  &\nodata&\nodata\\
4452.098 & [\ion{Fe}{2}]  & 7F  & 3.221e--14 &\nodata\\
4457.945 & [\ion{Fe}{2}]  & 6F  & 2.627e--14 &\nodata\\
4474.904 & [\ion{Fe}{2}]  & 7F  & 1.817e--14 &\nodata\\
4488.749 & [\ion{Fe}{2}]  & 6F  & 8.752e--15 &\nodata\\
4492.634 & [\ion{Fe}{2}]  & 6F  & 4.873e--15 &b\\
4509.602 & [\ion{Fe}{2}]  & 6F  & 7.680e--15 &e\\
4514.900 & [\ion{Fe}{2}]  & 6F  & 6.475e--15 &e\\
4639.667 & [\ion{Fe}{2}]  & 4F  & 1.049e--14 &f\\
4664.440 & [\ion{Fe}{2}]  & 4F  & 3.666e--15 &e\\
4728.068 & [\ion{Fe}{2}]  & 4F  & 2.069e--14 &\nodata\\
4774.718 & [\ion{Fe}{2}]  & 20F & 1.120e--14 &\nodata\\
4814.534 & [\ion{Fe}{2}]  & 20F & 4.495e--14 &\nodata\\
4861.325 & H$\beta$ & 1   &\nodata&\nodata\\
4874.485 & [\ion{Fe}{2}]  & 20F & 1.459e--14 &\nodata\\
\sidehead{\it inter-chip gap}
4973.388 & [\ion{Fe}{2}]  & 20F & 1.836e--14 &g\\
5005.512 & [\ion{Fe}{2}]  & 20F & 2.225e--14 &\nodata\\
5018.440 &  \ion{Fe}{2}   & 42  & 2.570e--14 &c,d\\
5020.233 & [\ion{Fe}{2}]  & 20F & 1.163e--14 &c,d\\ 
5041.024 &~ \ion{Si}{2}?  & 5   &\nodata&\nodata\\
5043.519 & [\ion{Fe}{2}]  & 20F &\nodata&\nodata\\
5055.984 &~ \ion{Si}{2}?  & 5   &\nodata&c\\
5056.317 &~ \ion{Si}{2}?  & 5   &\nodata&c\\
5158.001 & [\ion{Fe}{2}]  & 18F & 3.723e--14 &c,d\\ 
5158.777 & [\ion{Fe}{2}]  & 19F & 2.433e--14 &c,d\\ 
5163.951 & [\ion{Fe}{2}]  & 35F &\nodata&\nodata\\
5169.033 &  \ion{Fe}{2}   & 42  & 2.411e--14 &\nodata\\
5181.948 & [\ion{Fe}{2}]  & 18F & 6.598e--15 &\nodata\\
5184.788 & [\ion{Fe}{2}]  & 19F & 1.915e--15 &b\\
5197.902 &  [\ion{N}{1}]   & 1F  &\nodata&h\\
5220.059 & [\ion{Fe}{2}]  & 19F & 1.203e--14 &g\\
5261.621 & [\ion{Fe}{2}]  & 19F & 2.979e--14 &f\\
5268.874 & [\ion{Fe}{2}]  & 18F &\nodata&\nodata\\
5273.346 & [\ion{Fe}{2}]  & 18F & 5.160e--14 &f\\
5333.646 & [\ion{Fe}{2}]  & 19F & 2.506e--14 &f\\
5347.653 & [\ion{Fe}{2}]  & 18F &\nodata&c\\
5376.452 & [\ion{Fe}{2}]  & 19F & 2.388e--14 &\nodata\\
5477.241 & [\ion{Fe}{2}]  & 34F & 8.680e--15 &g\\
5527.609 & [\ion{Fe}{2}]  & 34F & 1.677e--14 &g\\
5577.339 &  [\ion{O}{1}]   & 3F  &\nodata&\nodata\\
5746.966 & [\ion{Fe}{2}]  & 34F & 1.329e--14 &\nodata\\
6300.304 &  [\ion{O}{1}]   & 1F  &\nodata&\nodata\\
6363.776 &  [\ion{O}{1}]   & 1F  &\nodata&\nodata\\
\sidehead{\it inter-chip gap}
7155.160 & [\ion{Fe}{2}]  & 14F & 3.861e--15 &\nodata\\
\enddata
\tablecomments{ The integrated
  line fluxes were measured on dereddened spectra and they are given in
  units of erg~cm$^{-2}$~s$^{-1}$.}  
\tablenotetext{a}{ ~multiplet 24 absent in NIST}
\tablenotetext{b}{ ~weak line; flux uncertain}
\tablenotetext{c}{ ~blend}
\tablenotetext{d}{ ~flux measurement obtained with a deblending
  procedure}   		
\tablenotetext{e}{ ~line with poorly define profile; flux uncertain}
\tablenotetext{f}{ ~difficult to define continuum; flux uncertain}
\tablenotetext{g}{ ~possible contribution to the flux from a molecular
  band} 		
\tablenotetext{h}{ ~blend with [\ion{N}{1}] 5200.257 \AA?}			         
 \end{deluxetable*}
\begin{deluxetable*}{c c c c c}
\tablewidth{0pt}
\tabletypesize{\small}
\tablecaption{P-Cyg and absorption lines\label{atom_ab_tab}}
\tablehead{
\colhead{$\lambda_{\rm lab}^{\rm air}$ (\AA)}& 
\colhead{Ion}&
\colhead{Multiplet}&
\colhead{Profile type}&
\colhead{Notes}}
\startdata
3933.663 & Ca II K  & 1   &absorption&a\\
3968.468 & Ca II H  & 1   &absorption&a\\
4254.332 &  Cr I    & 1   &absorption&b\\
4274.796 &  Cr I    & 1   &absorption&b\\
4289.716 &  Cr I    & 1   &absorption&b\\
4545.945 &  Cr I    & 10  &absorption&\nodata\\
4554.033 &  Ba II?  & 1   &absorption&c\\
4571.096 &  Mg I    & 1   & P-Cygni  &d\\	  
4580.043 &  Cr I    & 10  &absorption&\nodata\\
4613.357 &  Cr I    & 21  &absorption&\nodata\\
4616.120 &  Cr I    & 21  &absorption&\nodata\\
4646.148 &  Cr I    & 21  &absorption&e\\
4651.282 &  Cr I    & 21  &absorption&e\\
\sidehead{\it inter-chip gap}
5060.078 &  Fe I    & 1   &absorption&\nodata\\
5110.413 &  Fe I    & 1   &P-Cygni&f\\	
5127.681 &  Fe I?   & 1   &absorption&\nodata\\ 
5147.479 &  Ti I    & 4   & absorption&g\\
5152.185 &  Ti I    & 4   & absorption&g\\
5204.506 &  Cr I    & 7   &absorption&e\\
5206.038 &  Cr I    & 7   &absorption&e\\ 
5208.419 &  Cr I    & 7   &P-Cygni?&e\\
5247.049 &  Fe I?   & 1   &absorption&\nodata\\ 
5264.157 &  Cr I    & 18  &absorption&e\\
5265.723 &  Cr I    & 18  &absorption&e\\
5296.691 &  Cr I    & 18  &absorption&e\\ 
5298.277 &  Cr I    & 18  &P-Cygni&e\\ 
5300.744 &  Cr I    & 18  &absorption&\nodata\\
5345.801 &  Cr I    & 18  &absorption&e\\ 
5348.312 &  Cr I    & 18  &absorption&e\\ 
5394.676 &  Mn I    & 1   & P-Cygni&\nodata\\     
5409.772 &  Cr I    & 18  & P-Cygni&\nodata\\    
5426.237 &  Ti I    & 3   &absorption&g?\\
5432.546 &  Mn I    & 1   & P-Cygni&\nodata\\     
5535.481 &  Ba I    & 2   & P-Cygni&\nodata\\
5698.520 &   V I    & 35  &absorption&\nodata\\
5703.580 &   V I    & 35  &absorption&\nodata\\
5706.980 &   V I    & 35  &absorption&\nodata\\ 
5727.050 &   V I    & 35  &absorption&e\\
5727.650 &   V I    & 35  &absorption&e\\
5889.951 &Na I D$_2$& 1   &absorption&a\\
5895.924 &Na I D$_1$& 1   &absorption&a\\
6102.723 &  Ca I    & 3   &absorption&\nodata\\
6122.217 &  Ca I    & 3   &absorption&\nodata\\
\sidehead{\it inter-chip gap}
6551.676 &  Fe I    & 13  &absorption&\nodata\\
6556.062 &  Ti I?   & 102 &absorption&\nodata\\
6572.779 &  Ca I    & 1   & P-Cygni&\nodata\\ 
6707.761 &  Li I    & 1   &absorption&e,h\\	   
6707.912 &  Li I    & 1   &absorption&e,h\\	   
7364.097 &  Ti I?   & 97  &absorption&\nodata\\
7698.974 &  K I     & 1   & P-Cygni&\nodata\\  
7800.268 &  Rb I    & 1   & P-Cygni&\nodata\\ 
7947.603 &  Rb I    & 1   & ~P-Cygni?&\nodata\\
\enddata
\tablecomments{Photospheric signatures of the B3V companion are not included.}   
\tablenotetext{a}{ ~ISM feature}
\tablenotetext{b}{ ~weak line; flux uncertain}
\tablenotetext{c}{ ~\ion{Cs}{1} 4555.8 \AA\ resonance line?}
\tablenotetext{d}{ ~absorption component very weak}
\tablenotetext{e}{ ~blend}
\tablenotetext{f}{ ~emission component weak}
\tablenotetext{g}{ ~two separated components}				   
\tablenotetext{h}{ ~P-Cygni profile?}
 \end{deluxetable*}
\clearpage \LongTables   
\begin{deluxetable}{llcrrrcc}
\tablewidth{0pt}
\tabletypesize{\small}
\tablecaption{Identified molecular bands\label{TabMol}}
\tablehead{
\colhead{$\lambda_{\mathrm{obs}}$} & 
\colhead{$\lambda_{\mathrm{lab}}$} &
\colhead{Molecule}&
\colhead{Electronic}&
\colhead{Feature}&
\colhead{Vel.}&
\colhead{Ref.}&
\colhead{Notes}\\
\colhead{(\AA)} & 
\colhead{(\AA)} &
\colhead{}&
\colhead{transition}&
\colhead{}&
\colhead{(km\,s$^{-1}$)}&
\colhead{}&
}
\startdata
4420.5   & 4421.521 & TiO  & $\alpha$~~(4,0) & R$_{2}$    &      & 1 &\\ 
4469.4   & 4470.54  & AlO  & B$^{2}\Sigma^{+}$ -- X$^{2}\Sigma^{+}$~~(2,0) & R  &       & 2 &\\ 
4494.8   & 4494.04  & AlO  & B$^{2}\Sigma^{+}$ -- X$^{2}\Sigma^{+}$~~(3,1) & R  &       & 2 &\\ 
4517.5   & 4516.40  & AlO  & B$^{2}\Sigma^{+}$ -- X$^{2}\Sigma^{+}$~~(4,2) & R  &       & 2 &\\ 
4582.5   & 4584.056 & TiO  & $\alpha$~~(3,0) & R$_{2}$ &       & 1 &\\ 
4624.8   & 4626.08  & TiO  & $\alpha$~~(4,1) & R$_{2}$ &       & 1 &\\ 
4647.7   & 4648.23  & AlO  & B$^{2}\Sigma^{+}$ -- X$^{2}\Sigma^{+}$~~(1,0) & R  &       & 2 &\\
4671.7   & 4672.02  & AlO  & B$^{2}\Sigma^{+}$ -- X$^{2}\Sigma^{+}$~~(2,1) & R  &       & 2 &\\ 
4694.5   & 4694.62  & AlO  & B$^{2}\Sigma^{+}$ -- X$^{2}\Sigma^{+}$~~(3,2) & R  &       & 2 &\\ 
4715.4   & 4715.54  & AlO  & B$^{2}\Sigma^{+}$ -- X$^{2}\Sigma^{+}$~~(4,3) & R  &       & 2 &\\ 
4736.0   & 4735.82  & AlO  & B$^{2}\Sigma^{+}$ -- X$^{2}\Sigma^{+}$~~(5,4) & R  & & 2 &\\ 
4738.0   & 4737.460 & TiO  & c$^{1}\Phi$ -- X$^{3}\Delta$~~(0,0) & R$_{13}$  &    & 2 &\\ 
4754.4   & 4754.29  & AlO  & B$^{2}\Sigma^{+}$ -- X$^{2}\Sigma^{+}$~~(6,5)&R&+58 $\pm$ 10& 2 &\\ 
4757.9   & 4759.005 & $^{46}$Ti$^{16}$O  & $\alpha$~~(2,0) & R$_{2}$ &--67 $\pm$ ~6 & 3 &a\\ 
4759.1   & 4759.973 & $^{47}$Ti$^{16}$O  & $\alpha$~~(2,0) & R$_{2}$ &     & 3 &\\ 
\nodata  & 4760.902 & $^{48}$Ti$^{16}$O  & $\alpha$~~(2,0) & R$_{2}$ &     & 3 &\\ 
4803.0   & 4804.333 & TiO  & $\alpha$~~(3,1) & R$_{2}$ &       & 1 &b\\ 
4840.95  & 4842.27  & AlO  & B$^{2}\Sigma^{+}$ -- X$^{2}\Sigma^{+}$~~(0,0) & R$_{2}$&--82 $\pm$ 6 & 2 &\\ 
4952.6   & 4954.562 & TiO  & $\alpha$~~(1,0) & R$_{2}$ & --119 $\pm$ 6 & 1 &\\ 
4997.9   & 4999.129 & TiO  & $\alpha$~~(2,1)  & R$_{2}$ &             &  1 &\\ 
5044.9   & 5044.501 & TiO  & $\alpha$~~(3,2)  & R$_{2}$ &             &  4 &\\ 
5078.0   & 5079.36  & AlO  & B$^{2}\Sigma^{+}$ -- X$^{2}\Sigma^{+}$~~(0,1) & R &--82 $\pm$ ~6 & 2 &\\ 
5100.7   & 5102.13  & AlO  & B$^{2}\Sigma^{+}$ -- X$^{2}\Sigma^{+}$~~(1,2) & R &--82 $\pm$ ~6 & 2 &\\ 
5122.0   & 5123.33  & AlO  & B$^{2}\Sigma^{+}$ -- X$^{2}\Sigma^{+}$~~(2,3) & R &--82 $\pm$ ~6 & 5 &\\ 
5143.7   & 5142.89  & AlO  & B$^{2}\Sigma^{+}$ -- X$^{2}\Sigma^{+}$~~(3,4) & R & +48 $\pm$ 12 & 2 &\\ 
5161.8   & 5160.98  & AlO  & B$^{2}\Sigma^{+}$ -- X$^{2}\Sigma^{+}$~~(4,5) & R & +48 $\pm$ 12 & 2 &\\ 
5164.6   & 5166.753 & TiO  & $\alpha$~~(0,0)  & R$_{2}$ &--125 $\pm$ 12 & 1 &c\\ 
5227.4    & 5228.2   &  VO  & C$^{4}\Sigma^{-}$ -- X$^{4}\Sigma^{-}$~~(2,0) & R$_4$ &--45 $\pm$ 12 & 2 &\\ 
5276      & 5275.8   &  VO  & C$^{4}\Sigma^{-}$ -- X$^{4}\Sigma^{-}$~~(3,1)  &         &      & 2 &\\ 
5308.4    & 5307.337 & TiO  & $\alpha$~~(3,3)   & R$_{2}$ &     & 4 &\\ 
5312      & 5310.928 & TiO  & $\alpha$~~(3,3)   & R$_{3}$ &     & 4 &\\ 
5319.0    & 5319.746 & TiO  & $\gamma'$~~(3,0) & R$_{1}$ &      & 4 &d\\ 
5320.4    & 5321.111 & TiO  & $\gamma'$~~(3,0) & Q$_{1}$ &      & 4 &d\\ 
5338.6    & 5341.240 & TiO  & $\gamma'$~~(3,0)  & R$_{2}$ &     & 4 &d\\ 
5340.4    & 5343.223 & TiO  & $\gamma'$~~(3,0)  & Q$_{2}$ &     & 4 &d\\ 
5356.5    & 5355.975 & TiO  & $\alpha$~~(4,4)  & R$_{2}$ &      & 4 &\\ 
5361.5    & 5363.528 & TiO  & $\gamma'$~~(3,0)  & R$_{3}$ &     & 4 &d\\ 
5365.0    & 5367.167 & TiO  & $\gamma'$~~(3,0)  & Q$_{3}$ &     & 4 &d\\ 
\nodata   & 5376.83  & AlO  & B$^{2}\Sigma^{+}$ -- X$^{2}\Sigma^{+}$~~(2,4)  & R & & 2 &\\ 
\nodata   & 5411.960 & TiO & $\gamma'$~~(4,1) & R$_{3}$ &        & 4 &\\ 
\nodata   & 5415.639 & TiO & $\gamma'$~~(4,1) & Q$_{3}$ &        & 4 &\\ 
5447.2    & 5448.233 & TiO & $\alpha$~~(0,1) & R$_{2}$ & --57 $\pm$ 18 & 1 &\\ 
5467.9   & 5469.3   &  VO & C$^{4}\Sigma^{-}$ -- X$^{4}\Sigma^{-}$~~(1,0) & R$_4$ &--77 $\pm$ 12 & 2 &\\ 
5497.7   & 5496.742 & TiO & $\alpha$~~(1,2) & R$_{2}$ & +55 $\pm$ 12 & 1 &\\ 
5500.4   & 5499.741 & TiO & $\alpha$~~(1,2) & R$_{3}$ & +36 $\pm$ 12 & 1 &\\ 
\nodata  & 5517.3   & VO  & C$^{4}\Sigma^{-}$ -- X$^{4}\Sigma^{-}$~~(2,1) & R$_4$ &   & 2 &\\ 
\nodata  & 5546.041 & TiO & $\alpha$~~(2,3) & R$_{2}$  &            & 1 &\\ 
5562.1   & 5562.130 & TiO  & $\gamma'$~~(2,0) & $^S$R$_{21}$ &       & 5 &\\ 
5568.9   & 5569.089 & TiO  & $\gamma'$~~(2,0) & R$_{1}$ &            & 5 &\\ 
5591.4   & 5592.609 & TiO  & $\gamma'$~~(2,0) & R$_{2}$ &            & 5 &\\ 
5616.4   & 5616.792 & TiO  & $\gamma'$~~(2,0) & R$_{3}$ &            & 5 &\\ 
5640.3   & 5640.849 & TiO  & $\gamma'$~~(3,1) & R$_2$    &           & 5 &\\ 
5662.0   & 5662.223 & TiO  & $\beta$~~(2,2) & R        &           &  5 &\\ 
5662.8,
5665.0   & 5665.19  & TiO  & $\gamma'$~~(3,1) & R$_3$    &       & 5 &\\ 
5669.5   & 5669.966 & TiO  & $\gamma'$~~(3,1) & Q$_3$    &       &  5 &\\ 
5688.5   & 5689.640 & TiO  & $\gamma'$~~(4,2) & R$_2$    &       &  5 &\\ 
5691.3   & 5692.270 & TiO  & $\gamma'$~~(4,2) & Q$_2$    &       &  5 &\\ 
5713.8   & 5714.805 & TiO  & $\gamma'$~~(4,2) & R$_3$    &       &  5 &\\ 
5718.4   & 5719.223 & TiO  & $\gamma'$~~(4,2) & Q$_3$    &       &  5 &\\ 
5735.2   & 5736.703 & VO   & C$^{4}\Sigma^{-}-{\rm X}^{4}\Sigma^{-}$~~(0,0) & R$_4$ &--78 $\pm$ ~6 & 2 &\\ 
5759.4,
5760.4   & 5758.741 & TiO & $\alpha$~~(0,2) & R$_{2}$ &   &  3 &\\ 
5773.2   & 5772.74  & ScO  & A$^{2}\Pi_{3/2}$ -- X$^{2}\Sigma^{+}$~~(2,1) &$^{R}$Q$_{21}$&&2&e\\ 
\nodata  & 5811.60  & ScO  & A$^{2}\Pi_{1/2}$ -- X$^{2}\Sigma^{+}$~~(2,1) & Q$_{1}$      &&2&e\\ 
5811.0   & 5809.872 & TiO  & $\alpha$~~(1,3) & R$_{2}$   &       &  3 &\\ 
5827.7   & 5827.024 & TiO  & $\gamma'$~~(1,0) & R$_{31}$ &       &  3 &\\ 
5837.3   & 5839.005 & TiO  & $\gamma'$~~(1,0) & R$_{21}$ &--58 $\pm$ 10&  3 &\\ 
\nodata  & 5847.593 & TiO  & $\gamma'$~~(1,0) & R$_{1}$  &       &  3 &\\ 
5948.2   & 5947.680 & TiO  & $\gamma'$~~(2,1) & R$_{3}$  &       &  3 &\\ 
5956.2   & 5954.419 & TiO  & $\gamma'$~~(2,1) & Q$_{3}$  &       &  3 &\\ 
5972.5   & 5972.17  & YO   & A$^{2}\Pi_{3/2}-{\rm X}^{2}\Sigma^{+}$~~(0,0) & $^{R}$Q$_{21}$ &--8 $\pm$ ~5& 2 &\\ 
5978.5   & 5977.433 & TiO  & $\gamma$~~(3,0)  & R$_{3}$ &            & 3 &\\ 
5980.8   & \nodata  & YO   & A$^{2}\Pi_{3/2}-{\rm X}^{2}\Sigma^{+}$~~(0,0) &                &--8 $\pm$ ~5& 9 &\\
5988.8   & 5987.72  & YO   & A$^{2}\Pi_{3/2}-{\rm X}^{2}\Sigma^{+}$~~(1,1) & $^{R}$Q$_{21}$ &+58 $\pm$ ~5 & 2 &\\ 
5999.2   & 6001.151 & TiO  & $\gamma$~~(3,0)  & R$_{2}$ &            & 3 &\\
6027.8   & 6028.184 & TiO  & $\gamma$~~(3,0)  & R$_{1}$ &            & 3 &\\
6034.4   & 6036.17  & ScO  & A$^{2}\Pi$ -- X$^{2}\Sigma^{+}$~~(0,0)&$^{R}$Q$_{2G}+{^R}$R$_{2G}$&--55 $\pm$ 10 & 2 &\\
\nodata  & 6036.684 & TiO  & $\gamma$~~(4,1)  & R$_3$ &            & 3 &\\  
6061.9   & 6060.782 & TiO  & $\gamma$~~(4,1)  & R$_2$ &       +55 $\pm$ ~5 & 5 &\\  
6065.2   & 6064.31  & ScO  & A$^{2}\Pi$ -- X$^{2}\Sigma^{+}$~~(0,0) & $^{R}$R$_{1G}$ & +45 $\pm$ 10 & 2 &\\
\nodata  & 6072.65  & ScO  & A$^{2}\Pi$ -- X$^{2}\Sigma^{+}$~~(1,1) & $^{R}$Q$_{2G}$ &      & 2 &\\
6077.5   & 6079.30  & ScO  & A$^{2}\Pi$ -- X$^{2}\Sigma^{+}$~~(0,0) & $^{Q}$Q$_{1G}+{^Q}$R$_{1G}$ &--55 $\pm$ 10 & 2 &\\
\nodata  & 6088.245 & TiO  & $\gamma$~~~(4,1) & R$_1$ &            & 3 &\\  
6116.0   & 6115.97  & ScO  & A$^{2}\Pi$ -- X$^{2}\Sigma^{+}$~~(1,1)&$^{Q}$Q$_{1G}+^{Q}$R$_{1G}$  &  & 2 &\\ 
6130.8   & 6132.097 & YO   & A$^2\Pi_{1/2}$ -- X$^2\Sigma^{+}$~~(0,0) & $^{R}$Q$_{21}$  &    & 9 &\\
6135.0   & 6134.472 & TiO  & $\gamma'$~~(0,0) & $^{T}$R$_{31}$ &   & 3 &\\ 
6144.5   & \nodata  & TiO  & $\gamma'$~~(0,0) & $^{S}$Q$_{31}$ &       & &\\ 
6146.9   & 6148.68  & TiO  & $\gamma'$~~(0,0) & $^{S}$R$_{21}$ & --90 $\pm$ 10  & 6 &\\ 
6157.1   & 6158.52  & TiO  & $\gamma'$~~(0,0) &  R$_{1}$     &       & 6 &\\ 
6185.1   & 6186.32  & TiO  & $\gamma'$~~(0,0) &  R$_{2}$     &       & 6 &\\ 
\nodata  & 6189.65  & TiO  & $\gamma'$~~(0,0) &  Q$_{2}$     &       & 6 &\\ 
6213.4   & 6214.93  & TiO  & $\gamma'$~~(0,0) &  R$_{3}$     &       & 6 &\\ 
6220.9   & 6222.26  & TiO  & $\gamma'$~~(0,0) &  Q$_{3}$     & --70 $\pm$ 10 & 6 &\\ 
6269.2   & 6268.86  & TiO  & $\gamma'$~~(1,1) &  R$_{3}$     &  0 $\pm$ 15   & 5 &\\ 
6276.3   & 6276.03  & TiO  & $\gamma'$~~(1,1) &  Q$_{3}$     &  0 $\pm$ 15   & 5 &\\ 
6293.2   & 6294.80  & TiO  & $\gamma$~~(2,0) &  R$_{3}$      &               & 5 &\\ 
6319.4   & 6321.21  & TiO  & $\gamma$~~(2,0) &  R$_{2}$      & --77 $\pm$ ~5  & 5 &\\ 
6349.5   & 6351.29  & TiO  & $\gamma$~~(2,0) &  R$_{1}$      & --80 $\pm$ ~5  & 5 &\\ 
\nodata  & 6357.33  & TiO  & $\gamma$~~(3,1) & R$_{3}$       &               & 5 &\\
6384.5   & 6384.18  & TiO  & $\gamma$~~(3,1) & R$_{2}$       & +58 $\pm$ ~5    & 5 &\\ 
6410.2   & 6408.26  & ScO  & A$^2\Pi_{3/2}$ -- X$^2\Sigma^+$~~(0,1) & $^R$Q$_{2G}$+$^R$R$_{2G}$ &  &2&f\\
6415.1   & 6414.75  & TiO  & $\gamma$~~(3,1) & R$_{1}$ &+58 $\pm$ ~5 & 5 &\\ 
6459.6   & 6457.78  & ScO  & A$^2\Pi_{1/2}$ -- X$^2\Sigma^+$~~(0,1) & $^Q$Q$_{1G}$+$^Q$R$_{1G}$ &  &2&f\\ 
6448.9   & \nodata  & TiO  & $\gamma$~~(4,2) & R$_{2}$       &      &   &\\
6649.3   & 6651.271 & TiO  & $\gamma$~~(1,0) & R$_{3}$       &      & 1 &\\ 
6679.1   & 6680.796 & TiO  & $\gamma$~~(1,0) & R$_{2}$       &      & 1 &\\ 
6712.5   & 6714.477 & TiO  & $\gamma$~~(1,0) & R$_{1}$       &      & 1 &\\ 
\nodata  & 6717.599 & TiO  & $\gamma$~~(2,1) & R$_{3}$       &      & 1 &\\ 
6747.0   & 6747.613 & TiO  & $\gamma$~~(2,1) & R$_{2}$       &      & 1 &\\ 
6781.1   & 6781.815 & TiO  & $\gamma$~~(2,1) & R$_{1}$       &      & 1 &\\ 
\nodata  & 6784.560 & TiO  & $\gamma$~~(3,2) & R$_{3}$       &      & 1 &\\ 
6815.9   & 6815.139 & TiO  & $\gamma$~~(3,2) & R$_{2}$       &      & 1 &\\ 
6832.2   & 6832.596 & TiO  & b$^{1}\Pi$ -- X$^{3}\Delta$~~\phantom{(2,0)} & R  &--82 $\pm$ ~8 & 3&g\\ 
6838.9   & 6838     & TiO  & b$^{1}\Pi$ -- X$^{3}\Delta$~~\phantom{(2,0)} &  Q,P    &         & 3 &\\ 
6849.9   & 6849.927 & TiO  & $\gamma$~~(3,2) & R$_{1}$       &      & 1 &\\ 
6884.3   & 6885.25  &  VO  & B$^{4}\Pi$ -- X$^{4}\Sigma^{-}$~~(2,0) &$^T$R$_{42}$  &      & 2 &\\ 
6892.0   & 6893.86  &  VO  & B$^{4}\Pi$ -- X$^{4}\Sigma^{-}$~~(2,0) &$^S$Q$_{42}$  &      & 2 &\\ 
6935.8   & 6937.    &  VO  & B$^{4}\Pi$ -- X$^{4}\Sigma^{-}$~~(2,0) &  &      & &\\
6950.9   & 6952.39  &  VO  & B$^{4}\Pi$ -- X$^{4}\Sigma^{-}$~~(2,0)  &  &      & &\\
6977.9   & 6978.51  &  VO  & B$^{4}\Pi$ -- X$^{4}\Sigma^{-}$~~(2,0) &  &      & &\\
6986.6   & 6985.59  &  VO  & B$^{4}\Pi$ -- X$^{4}\Sigma^{-}$~~(2,0) &  &      & &\\ 
7003.5   & 7003.40  & TiO  & $\gamma$~~(0,0) & $^{S}$R$_{32}$&& 4 &\\ 
7037.8   & 7038.28  & TiO  & $\gamma$~~(0,0) & $^{S}$R$_{21}$&& 4 &\\ 
\nodata  & 7054.256 & TiO  & $\gamma$~~(0,0) & R$_{3}$ &      & 1 &\\ 
7085.5   & 7087.566 & TiO  & $\gamma$~~(0,0) & R$_{2}$ &      & 1 &\\ 
\nodata  & 7124.930 & TiO  & $\gamma$~~(1,1) & R$_{3}$ &      & 1 &\\ 
7123.4   & 7125.510 & TiO  & $\gamma$~~(0,0) & R$_{1}$ &      & 1 &\\ 
\nodata  & 7158.850 & TiO  & $\gamma$~~(1,1) & R$_{2}$ &      & 1 &\\ 
7231.8   & 7230.774 & TiO  & $\gamma$~~(2,2) & R$_{2}$ &      & 3 &\\ 
7270.9   & 7269.985 & TiO  & $\gamma$~~(2,2) & R$_{1}$ &      & 3 &\\ 
7332.5   & 7333.93  &  VO  & B$^{4}\Pi$ -- X$^{4}\Sigma^{-}$ (1,0) & $^T$R$_{42}$ &--60 $\pm$ ~8 & 7&h\\ 
7371.5   & 7372.70  &  VO  & B$^{4}\Pi$ -- X$^{4}\Sigma^{-}$ (1,0) & $^S$Q$_{31}$ &--60 $\pm$ ~8 & 7&h\\ 
\nodata  & 7381.70  &  VO  & B$^{4}\Pi$ -- X$^{4}\Sigma^{-}$ (1,0) & $^R$P$_{31}$ &	& 7 &\\ 
\nodata  & 7393.34  &  VO  & B$^{4}\Pi$ -- X$^{4}\Sigma^{-}$ (1,0) & $^S$R$_{21}$ &	& 7 &\\ 
\nodata  & 7405     &  VO  & B$^{4}\Pi$ -- X$^{4}\Sigma^{-}$ (1,0) &		   &	& 7 &\\ 
\nodata  & 7416.13  &  VO  & B$^{4}\Pi$ -- X$^{4}\Sigma^{-}$ (1,0) & $^Q$P$_{21}$ &	& 7 &\\ 
\nodata  & 7433.79  &  VO  & B$^{4}\Pi$ -- X$^{4}\Sigma^{-}$ (1,0) & R$_{1}$	   &	& 7 &\\ 
\nodata  & 7442.9   &  VO  & B$^{4}\Pi$ -- X$^{4}\Sigma^{-}$ (1,0) &		   &	& 7 &\\ 
7453.8   & 7455.2   &  VO  & B$^{4}\Pi$ -- X$^{4}\Sigma^{-}$ (1,0) &              &    & 7 &\\ 
\nodata  & 7474     &  VO  & B$^{4}\Pi$ -- X$^{4}\Sigma^{-}$ (2,1) &              &+58 $\pm$ ~8& 7&h\\ 
7589.8   & 7589.284 & TiO  & $\gamma$~~(0,1) & R$_{3}$       &      & 1 &\\ 
7627.5   & 7627.799 & TiO  & $\gamma$~~(0,1) & R$_{2}$       &      & 1 &\\ 
7704.2   & 7704.997 & TiO  & $\gamma$~~(1,2) & R$_{2}$       &      & 1 &\\ 
7744.4   & 7743.056 & TiO  & $\gamma$~~(2,3) & R$_{3}$       &      & 1 &\\ 
7750.9   & 7749.487 & TiO  & $\gamma$~~(1,2) & R$_{1}$       &      & 1 &\\ 
7783.8   & 7782.854 & TiO  & $\gamma$~~(2,3) & R$_{2}$       &      & 1 &\\ 
7863.20  & 7865.29  &  VO  & B$^{4}\Pi$ -- X$^{4}\Sigma^{-}$~~(0,0) & $^{S}$Q$_{42}$ &--80 $\pm$ ~4 &  8 &h\\ 
7894.20  & 7896.38  &  VO  & B$^{4}\Pi$ -- X$^{4}\Sigma^{-}$~~(0,0) & $^S$Q$_{31}$   &--80 $\pm$ ~4 &  8 &h\\ 
7905.72  & 7907.80  &  VO  & B$^{4}\Pi$ -- X$^{4}\Sigma^{-}$~~(0,0) & $^R$P$_{31}$   &--80 $\pm$ ~4 &  8 &h\\ 
\nodata  & 7919.12  &  VO  & B$^{4}\Pi$ -- X$^{4}\Sigma^{-}$~~(0,0) & $^S$R$_{21}$   &             &  8 &\\ 
7937.55  & 7939.62  &  VO  & B$^{4}\Pi$ -- X$^{4}\Sigma^{-}$~~(0,0) & R$_{2}$        &--80 $\pm$ ~4 &  8 &h\\ 
\nodata  & 7950.29  &  VO  & B$^{4}\Pi$ -- X$^{4}\Sigma^{-}$~~(0,0) & $^{Q}$R$_{23}$ &             &  8 &\\ 
\enddata
\tablerefs{(1)~\cite{PHI73}; (2)~\cite{KOP74}; (3)~\cite{SCH98}; (4)~\cite{PLE98}; 
         (5)~\cite{JOR94}; (6)~\cite{HGM79}; (7)~\cite{CHE94}; (8)~\cite{ABB95};
         (9)~\cite{BER83}.}
\tablenotetext{a}{ After correcting the laboratory wavelength
                  from \cite{PHI73} for the isotopic shift, as
                  found from data in \cite{SCH98}, one gets for the
                  isotopomer $^{46}$TiO the predicted
                  stationary wavelength of 4759.205\,\AA\ and
                  the velocity of --80\,km\,s$^{-1}$.}	
\tablenotetext{b}{ The bandhead is strongly contaminated by the rotational structure of 
                   the $\alpha$ (2,0) band of TiO in the range 4793--4803\,\AA.}
\tablenotetext{c}{ The bandhead is blended with two \ion{Fe}{1} absorption lines 
                  at 5166.286\,\AA\ and 5168.901\,\AA. However, the
                  blue edge of the blend should be formed by molecular absorption. 
		  The velocity given in the table is for the
                  $^{46}$TiO isotopomer (laboratory wavelength
                  concluded from \citealp{SCH98}).}			
\tablenotetext{d}{ All databases give different laboratory wavelengths.}   
\tablenotetext{e}{ The band is probably contaminated
                  by the A$^2\Pi_{1/2}$--X$^2\Sigma^+$ (1,0) Q$_1$ feature of ScO.}
\tablenotetext{f}{ In emission.}
\tablenotetext{g}{ Velocity determined in simulations, see Fig.~\ref{TiO_f_fig} for details.}
\tablenotetext{h}{ In the table the velocity of the blueshifted component of VO is given. The
                  fit to all features of the VO (1,0) and (0,0) bands indicates a velocity of --2\,km\,s$^{-1}$.}  
\end{deluxetable}

\end{document}